\documentclass[12pt,a4paper]{article}
\usepackage[margin=2cm]{geometry}
\usepackage[margin=5pt,labelfont=bf]{caption}
\usepackage{graphicx,amsmath,amssymb,natbib,setspace,
  xspace,color,appendix,authblk}
\usepackage{subfig}
\definecolor{darkblue}{rgb}{0.0,0.0,0.6}
\usepackage[colorlinks=true,breaklinks=true,linkcolor=darkblue,
urlcolor=darkblue,anchorcolor=darkblue,citecolor=darkblue]{hyperref}
\usepackage{lineno}
\usepackage{epstopdf}

\usepackage[T1]{fontenc}
\usepackage{multirow} %

\newcommand{\diff}[2]{\dfrac{\text{d}{#1}}{\text{d}{#2}}}
\newcommand{\pdiff}[2]{\dfrac{\partial{#1}}{\partial{#2}}}
\newcommand{\cotwo}{\ensuremath{\text{CO}_2}}

\newcommand{\eg}{\textit{e.g.,}}
\newcommand{\ie}{\textit{i.e.,}}

\newcommand{\gT}{\ensuremath{\gamma_{\textrm{T}} }}
\newcommand{\gMOR}{\ensuremath{\gamma_{\textrm{MOR}} }}
\newcommand{\gSAV}{\ensuremath{\gamma_{\textrm{SAV}} }}
\renewcommand{\deg}{\ensuremath{^\circ}} %
\newcommand{\dOxy}{\ensuremath{\delta^{18}\textrm{O}}}

\title{Modelling Coupled Oscillations of  Volcanic CO$_{2}$ Emissions and Glacial Cycles}

\author[$1$]{Jonathan M.A.~Burley\thanks{jonathan.burley@earth.ox.ac.uk}} 
\author[$2$]{Peter Huybers} 
\author[$1$]{Richard F.~Katz} 
\affil[$1$]{\small Department of Earth Sciences, University of
  Oxford, UK}
\affil[$2$]{\small Department of Earth and Planetary Sciences, Harvard
University, Cambridge, USA}

\begin{document}
\maketitle
\section{Introduction} \label{sec:intro}
Following the mid-Pleistocene transition, glacial cycles changed from
$40$~kyr cycles to longer $80$ or $120$~kyr
cycles\citep{Lisiecki:2005iu,Elderfield:2012cf}.  The 40~kyr glacial
cycles are broadly accepted as being driven by cyclical changes in
Earth's orbital parameters and the consequent insolation changes ---
Milankovitch cycles.  However, Milankovitch forcing does not readily
explain the $>40$~kyr glacial cycles that occur after the
mid-Pleistocene transition.  These $>40$~kyr cycles therefore
require that internal dynamics in the Earth system create a
glacial response that is not linearly related to insolation \citep{Tziperman:2006he}.

Any proposed mechanism to extend glacial cycles' periods beyond
40~kyrs must give the Earth's climate system a memory on the order of
10s-of-kyrs, creating either a response that counteracts the 40~kyr
Milankovitch forcing (allowing the Earth to `skip' beats in the 40~kyr
forcing) or a climate state with sufficient inertia --- low climate
sensitivity --- that it is not affected by 40~kyr obliquity forcing
\citep{Imbrie:1980bg}.  The atmosphere/ocean has typical adjustment
timescales on the order of $1000$~years, thus oceanic theories for
glacial cycles rely on other, long-timescale processes (eg. weathering
\citep{Toggweiler:2008ce}) to trigger arbitrary rules-based switches
in the oceanic carbon system at tens-of thousands-of-years intervals.
Hence, it is difficult to envision how the ocean and atmosphere system
could disrupt 40~kyr glacial cycles with a counteraction or inertia
response; other mechanisms must be involved.

Hypothesised mechanisms of climate-inertia include: Antarctic ice
sheets limiting deepwater ventilation \citep{Ferrari:2014gr},
erosion of regolith to high-friction bedrock creating a thicker
Laurentide icesheet \citep{Clark:1998fx}, ice-sheet calving
instabilities \citep{Pollard:1983tp}, and sea-ice limiting
precipitation over ice sheets \citep{Gildor:2000vi}. However, none of
these are universally accepted.
More recently, \cite{AbeOuchi:2013hm} proposed a model of $\sim$100~kyr glacial
cycles for the past 400~kyrs.  They modelled a 3D ice sheet forced by
insolation and a prescribed \cotwo{} timeseries, using parameterised
changes to temperature and precipitation derived from snapshots of a
GCM (General Circulation Model).  The reason for their $\sim$100~kyr
cycles is the climate-inertia of the Laurentide ice sheet: when the ice sheet
is small it grows or remains stable in response to orbit-induced and
\cotwo{}-induced climate perturbations, however at a larger size the
Laurentide becomes unstable to such perturbations and will rapidly
retreat in response to a warming event.

The large Laurentide ice-sheet's instability to warming perturbations
is due to isostatic lithospheric adjustments forming a depression underneath an
old ice sheet \citep{Oerlemans:1980ic,Pollard:1982wd}.  The retreat of
the ice sheet is also a retreat downslope (in the isostatic
depression), continually exposing the ice sheet to warmer air, a
positive feedback.

The \cite{AbeOuchi:2013hm} model is not unique in producing 100~kyr
cycles, \cite{Ganopolski:2011dq} manage the same in a slightly lower
complexity model with an instability to warming perturbations derived
from increased dust feedback when the Laurentide moves far enough south
to encounter sediment-rich locations.  \cite{Ganopolski:2011dq} state
that any non-linear feedback on ice retreat could likely produce the
same behaviour (although isostatic lithospheric adjustments are not
sufficient in their model).

But, even in the framework of 100~kyr ice-sheet hysteresis, an
explanation of late-Pleistocene glacial cycles must also explain why
\cotwo{} minima (of the appropriate magnitude) occur on $100$~kyr
periods.  The \cite{AbeOuchi:2013hm} model calculates approximate
$100$~kyr cycles when \cotwo{} is fixed at $220$~ppmv.  This fixed,
glacial \cotwo{} value makes the Laurentide ice sheet unstable to
orbital variations at $\sim$90~msle (metres sea-level equivalent)
global ice volume.  Fixed atmospheric \cotwo{} values significantly
above or below $220$~ppmv prevent the $\sim$100~kyr cycles from
emerging.  

Furthermore, whilst the \cite{AbeOuchi:2013hm} fixed-\cotwo{} scenario
has a predominant 100~kyr cycle, the resulting sea-level timeseries
has departures from the geological record that are not present when
prescribing \cotwo{}: \textit{i}) the power spectrum's 23~kyr and
40~kyr signals have similar strength, rather than a 1:2
ratio. \textit{ii}) the last deglaciation and MIS11 deglaciation are
small, leaving large ice-sheets at peak `interglacial'.  Thus, even a
carefully selected fixed \cotwo{} value does not allow a model to
replicate glacial behaviour; suggesting there is a need to incorporate
a dynamic \cotwo{} response to fully understand glacial cycles.

For over a century \citep{Arrhenius:1896wr}, it has been known that
the $\sim$$2,200$ Gt of \cotwo{} in the atmosphere is connected to
much larger carbon reservoirs --- there are $147,000$~Gt\cotwo{} in
oceans and ocean sediments, and $9,200$~Gt\cotwo{} in the biosphere
and soils, and $200,000,000$~Gt\cotwo{} in the mantle
\citep{IPCC:2013_PhysSci,Dasgupta:2010fw} --- and that an imblance in
fluxes between them could alter atmospheric \cotwo{}
concentration. 

Despite this, exact mechanisms behind $\sim$100~kyr variations in
atmospheric \cotwo{} concentration are unknown. Several theories based
on ocean--atmosphere \cotwo{} partitioning exist, and can generate the
total atmospheric \cotwo{} change (although this could be acheived
without oceanic partitioning
\citep{Crowley:1995fc,Adams:1998vw,Ciais:2012kh}); however, they do
not make satisfactory dynamic predictions for the timing and magnitude
of the observed atmospheric \cotwo{} record, nor the oceanic carbonate
record \citep{Broecker:2015kw}.

The line of argument for ocean--atmosphere \cotwo{} partitioning
theories, simplified somewhat to summarise here, involves changing
surface ocean and deep water exchange locations and volumes (and
consequent changes in ocean carbonate chemistry). These can
cumulatively change atmospheric \cotwo{} concentration by roughly
80~ppmv, be it by reorganising ocean currents
\citep{Toggweiler:1999wj}, ice sheets altering ocean
ventilation \citep{Ferrari:2014gr}, changing the biological pump
via nutrient control \citep{Sigman:2010bz}, or southern ocean wind
stress \citep{Franois:1997jy}.  These theories share similar features:
from interglacial conditions, a reduction in planetary temperature
triggers a change in an ocean-relevant process; consequently, altered
ocean behaviours sequester \cotwo{} in the deep ocean, reducing
atmospheric \cotwo{} concentration and acting as a positive feedback
to the initial temperature change.  However, predicting these trigger
points and calculating appropriate atmospheric \cotwo{} reduction
rates (rather than just total \cotwo{} reduction) over a full glacial
cycle remains infeasible.

\cite{Broecker:2015kw} notes that ocean-only mechanisms for the
glacial \cotwo{} cycle necessitate a deep-sea carbonate preservation
event during deglaciation. However, no such event is seen in ocean
sediment records.

Reconciling oceanic observations with theory would be possible with a
variable \cotwo{} flux into the ocean-atmosphere reservoir ---
\cite{Broecker:2015kw} suggest that a previously hypothesised,
glacially-induced variability in volcanic emissions would be suitable.

We have discussed two features of the glacial \cotwo{} record that
volcanic \cotwo{} emissions could help explain: first, the long drawdown
of \cotwo{} over $\sim$100~ kyrs (ie. Earth's climate
system has a memory on the order of 10s-of-kyr), and second, increased
\cotwo{} during deglaciation. How can volcanic \cotwo{} emissions
perform these roles?

Recent work suggests volcanic \cotwo{} emissions change in response to
glacial cycles \citep{Huybers:2009hn,Tolstoy:2015kq,Burley:2015ic}:
subaerial volcanic \cotwo{} emissions respond to glaciation within a
few thousand years
\citep{Huybers:2009hn,Kutterolf:2013dd,Rawson:2015ug}, and mid-ocean
ridge (MOR) \cotwo{} emissions respond to changing sea level with a
10s-of-kyrs delay \citep{Burley:2015ic}.  This MOR delay occurs,
according to \cite{Burley:2015ic}, because changing sea-level causes a
\cotwo{} anomaly in mantle melt at about $60$~km below the MOR, where
hydrous melting abruptly becomes silicate melting.  This \cotwo{}
anomaly subsequently takes tens of thousands of years to be carried to
the MOR axis by the melt transport.

\begin{figure}[ht]
  \centering
  \includegraphics[width=12cm]{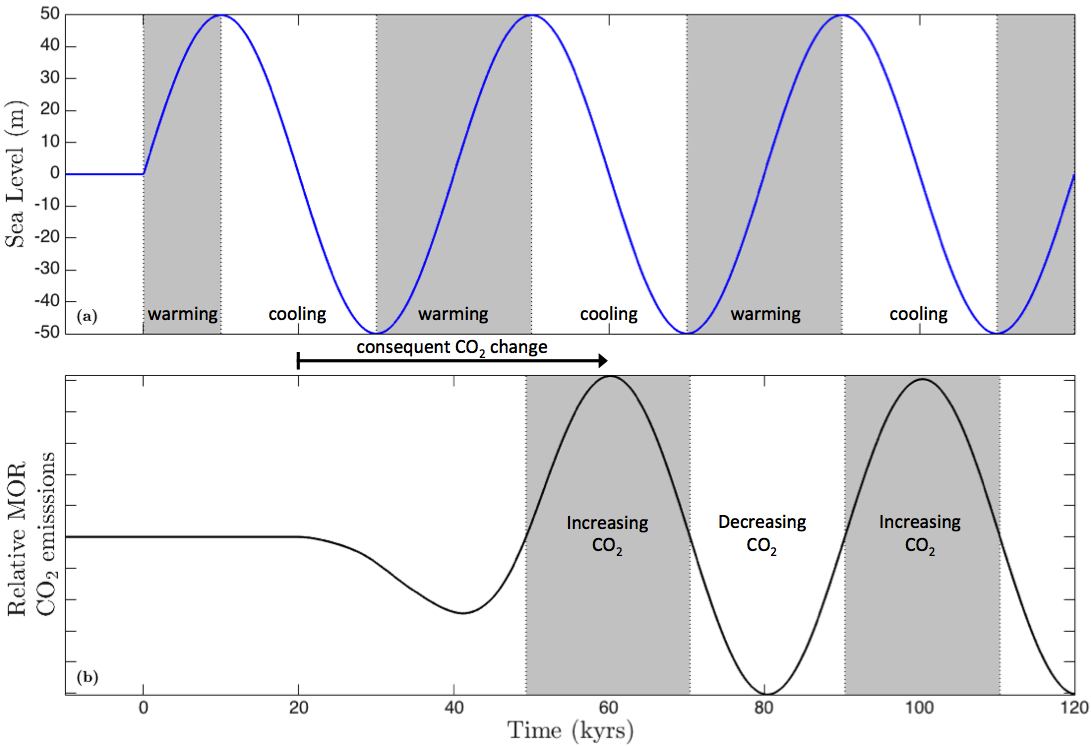}
  \caption{ \textbf{(a)} Sinusoidal sea level change and \textbf{(b}) consequent global
    MOR \cotwo{} emissions for a 40~kyr lag (using
    \cite{Burley:2015ic} calculations the emissions are, roughly,
    proportional to minus the rate of change of sea-level offset by
    the lag time).  Sea level is treated as an
    independent input in this figure.  The grey shading denotes a
    warming climate in panel (a) and increasing \cotwo{} in panel
    (b). MOR \cotwo{} emissions oppose the SL change when grey regions
    overlap white regions between the two panels; MOR lag of 30--50
    kyrs gives regimes that oppose 40~kyr sea level more than they
    reinforce it.}
  \label{fig:MORC_didactic}
\end{figure}

Conceptually, as shown in figure~\ref{fig:MORC_didactic}, MOR \cotwo{}
emissions that lag sea-level by 30--50~kyrs would act to create high
atmospheric \cotwo{} concentration in periods of low insolation.  This
lagged MOR \cotwo{} emissions response gives the Earth system a memory
on the 50~kyr timescale that could act to drive glacial cycles from
40~kyr cycles to a multiple of this period.  If so, such glacials
would have sawtooth profile; entering a glacial under insolation
forcing, with a hiatus in ice sheet growth as increasing insolation and
low \cotwo{} concentration counteract each other, followed by a deeper
glacial as insolation reduces, then a large deglaciation as both
insolation and \cotwo{} increase.

By contrast, variable subaerial volcanic \cotwo{} emissions (with a
few thousand year lag) are unlikely to change the period of glacial
cycles, acting instead as a positive feedback on changes in ice volume
\citep{Huybers:2009hn} \eg{} increasing \cotwo{} during deglaciation.

The lagged MOR \cotwo{} response's effect on glacial cycles was first
investigated in \cite{Huybers:2017vz} using coupled differential
equations to parameterize global ice volume, average temperature, and
atmospheric \cotwo{} concentration.  Ice volume changes at a rate
proportional to both the current temperature and ice volume to the
third power (the latter gives a maximum and minimum bounding ice
volume). Temperature varies according to insolation, temperature, and
atmospheric \cotwo{} concentration.  Atmospheric \cotwo{}
concentration varies according to average temperature, subaerial
volcanism and MOR volcanism (based on \cite{Burley:2015ic}
calculations).  These equations represent a coupled, non-linear
oscillator, and generate glacial cycles at a multiple of the obliquity
period.

These results are intriguing, however there are limits in their
physical representation.  For instance, they have: \textit{1)} an insolation
forcing timeseries with no seasonal or spatial component;  \textit{2)}
a negative ice feedback proportional to the volume of the ice sheet
cubed, inducing symmetrical variability rather than sawtooth behaviour;
\textit{3)} no isostatic lithospheric response to the ice sheet.
These simplifications remove potentially important
physical mechanisms from the model glacial system.

These results are intriguing, however a more complete representation
would allow more detailed consideration of the key physics.
The present study builds on \cite{Huybers:2017vz} by extending the
modelling framework to a low complexity earth system model.  We ask:
what properties the volcanic \cotwo{} response to glacial cycles must
have to alter the period of a glacial cycle?

We extend a simplified climate model from \cite{Huybers:2008fg} which
focused on accurate treatment of orbital forcing, using an Energy
Balance Model (EBM) coupled with an ice sheet model.  The EBM
calculates daily insolation to resolve the counteracting effects of
orbital precession on ice sheets: hotter but shorter summers. The
\cite{Huybers:2008fg} model demonstrates 40~kyr glacial cycles in
response to insolation forcing. To maintain their focus on orbital
effects, they did not consider the radiative effects of varying
atmospheric \cotwo{} and water vapour; they assumed an atmosphere of
constant composition.  From that framework we extend to a system of
three component models: energy balance, ice sheet growth, and \cotwo{}
concentration in the atmosphere.  Our model does not aim to be perfect
representation of the climate; rather it focuses on approximating key
features and feedbacks such that we can calculate Earth's glacial
state over several glacial cycles.

Previous models of glacial cycles have ranged from simple, abstracted
systems \citep{Imbrie:1980bg} to detailed representations of ice
sheets and climate physics \citep{AbeOuchi:2013hm} --- our model
complexity is partway along this spectrum, considering the essential
physics acting on a  pseudo-2D system.  However, even
\cite{AbeOuchi:2013hm} omit the carbon cycle, using imposed \cotwo{}
concentrations rather than a dynamic system. No model has yet fully
coupled an explicit representation of the solid-earth carbon cycle to
physical representations of the Earth's climate.  This work presents
such a fully coupled model using a low-complexity physical
representation.  The full insolation forcing is used to drive an Earth
system response in \cotwo{} concentration, temperature, and ice sheet
configuration.

We will show that this model,  when forced by the observed \cotwo{}
record, calculates sea-level timeseries that closely match the
historical record.  When \cotwo{} evolves freely, the model has no $\sim$100~kyr
sea level variability until we include the lagged MOR \cotwo{}
feedback; it is necessary to have a \cotwo{} feedback process with a
period similar to or greater than the default 40~kyr glacial cycle in
order to disrupt that cycle.
We will show that the variation in MOR \cotwo{} emissions has the
potential to generate sawtooth glacials.
The importance of volcanism in glacial cycles depends on both the
percentage variations in volcanic emissions during glacial cycles and
the background volcanic \cotwo{} emissions rate.  There are
uncertainties in both these quantities for MOR and subaerial systems.
This uncertainty guides the modelling choices in made here. Rather
than attempt a single exact estimate of global volcanic effects, we
instead consider a range of volcanic effects. We define the threshold
at which volcanism changes the pacing of glacial cycles, and compare
this to estimates of these volcanic quantities.  If the threshold
values are orders of magnitude outside of estimates of these
quantities, it would be strong evidence that volcanic \cotwo{}
variability is not a important mechanism in glacial cycles.  Our model
scales linearly with changes in baseline volcanic emissions and
volcanic variability, so our results can be readily reinterpreted if
such estimates are updated.

As mentioned above, our preference in this work is to consider the
Earth's early-Pleistocene glacials as 40~kyr cycles with an internal
Earth system feedback that locked the Earth into a 100~kyr mode after
the mid-Pleistocene transition.  The model system is agnostic about
the orbital forcing responsible for this; our forcing includes the
full insolation distribution in precession, obliquity, and
eccentricity.  Whilst it would be possible to parse the relative
influence of obliquity and precession index forcing, it is not needed
in the present context.

Section~\ref{sec:method} introduces the three component models used
to generate our results and discusses their coupling.
Section~\ref{sec:results} contains demonstrations of conceptually
important model behaviour and the key model
results: Section~\ref{sec:mor-co_2-response} discusses how sea level
period controls the maximum atmospheric \cotwo{} anomaly induced by
MOR volcanoes. Section~\ref{sec:IceCoreRep} demonstates our model's
agreement with historical sea-level data when forced by the ice core
\cotwo{} record. Section~\ref{sec:varying-mor-lag} investigates the
climate effects of different MOR lag times under simplified orbital
forcing and discusses the importance of different timescale \cotwo{}
feedbacks.
Section~\ref{sec:full-model-behaviour} demonstrates model behaviour
for a range of potential \cotwo{} feedbacks, showing the circumstances
under which $\sim$100~kyr cycles
occur. 
Section~\ref{sec:discussion} discusses the significance of assumptions
and simplifications made in the model and the meaning of our results.
Section~\ref{sec:conclusion} summarises our findings and offers some
conclusions.

\section{Method} \label{sec:method}
The research question we ask, regarding the pacing of glacial cycles,
requires that the model must run
for 100's of thousands of years.
To be capable of this, the model must use a reduced complexity
representation of the climate system.
The model treats the Earth's climate as a record of ice sheet volume
(equivalently, sea level), temperature, and the \cotwo{} concentration
in the atmosphere. We consider 2D models of ice and temperature,
modelling a line from the equator to north pole.

\begin{figure}[ht]
  \centering
  \includegraphics[width=7cm]{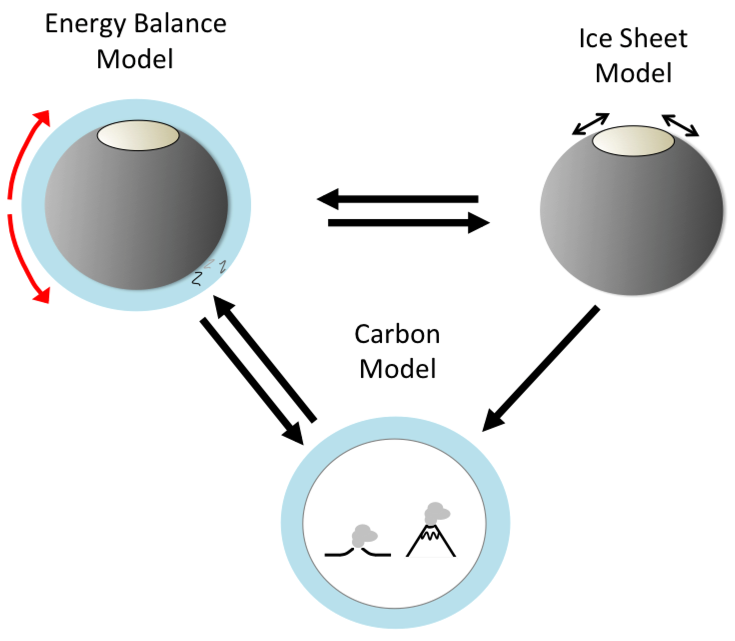}
  \caption{Component models and their interactions. Black arrows
    between models show information flow.  Energy balance (\ie{}
    temperature) is affected by carbon concentration in the atmosphere
    and the extent of the ice sheet.  The ice sheet is affected by
    temperature only.  Carbon concentration in the atmosphere is
    affected by the rate of change of temperature and sea level
    (determined by the ice sheet).}
  \label{fig:ModellingHat}
\end{figure}

Independent variables are time $t$ and latitude $\phi$.
Temperature $T$ is a function of $t,\phi$,  changing due to insolation
$S$, ice (\ie{} surface albedo),
atmospheric \cotwo{} concentration, current temperature (controls
longwave infrared emissions), and the temperature gradient with
latitude.
Ice sheet thickness $h$ is a function of $t,\phi$.  It changes as ice
flows under its own weight and accumulates/melts according to local
temperature.  Integrating $h$ over latitude $\phi$  --- with an assumed ice
sheet width --- gives total ice volume $V$.
The \cotwo{} concentration in the atmosphere $C$ is a function of $t$,
varying in response to three processes: $T$-dependent changes in the
surface system (\ie{} atmosphere, biosphere, and ocean) partitioning
of \cotwo{}, $V$-dependent changes in subaerial volcanism (SAV), and
$V$-dependent changes in mid-ocean ridge (MOR) volcanism.  The
dependencies of these components are shown graphically in
figure~\ref{fig:ModellingHat}.

These components are described by the following differential equations
\begin{align}
  \pdiff{T(t,\phi)}{t} &= f_T \left( S,h,C, T , \pdiff{T}{\phi}\right) \;\;, \label{eq:TopLvl_Temp}\\
  \pdiff{h(t,\phi)}{t} &= f_h \left(\pdiff{h}{\phi}, T  \right)\;\;, \label{eq:TopLvl_Ice}\\ %
  \diff{C(t)}{t} &= f_C  \left( \pdiff{T}{t}  , \diff{V}{t} \right)
    \;\;, \label{eq:TopLvl_Catm}
\end{align}
where functions $f_i$ determine the rate of change of variable
$i$.
 The system of equations
\eqref{eq:TopLvl_Temp}--\eqref{eq:TopLvl_Catm} is driven by variation
in insolation, $S$, computed using \cite{Berger:1991fm}. All other
variables evolve in response to the internal state of the model.
Conceptually, this matches the Earth system:  internal dynamics
affected by the external driving force of variable insolation.

Having discussed the way these component models will be linked,
  we now describe each model in detail.

\subsection{Energy Balance Model} \label{sec:energy-balance-model}
  To calculate planetary temperature and the annual ice
  accumulation/melting we use an Energy Balance Model (EBM) based on
  \cite{Huybers:2008fg}.  This model calculates insolation (and
  consequent temperature changes) at daily intervals, thus explicitly
  modelling the seasonal cycle and its effect on ice sheet
  accumulation/melting. Importantly, this includes the counteracting
  effects of orbital precession on ice sheets: hotter but shorter
  summers.

  The EBM is fully detailed in \cite{Huybers:2008fg}, here we will
  briefly cover the overall model and explain our method for including
  radiative forcings to represent \cotwo{}, water vapour, lapse rate,
  and cloud effects.

The EBM tracks energy in the atmosphere, ground surface, and
 subsurface; this is encompassed in:
\begin{align}
c_a \pdiff{T_a}{t} = S_a + I_a + F_s + D_a \;\;,\label{eq:EBM_atm}\\
c_s \pdiff{T_s}{t} = S_s + I_s - F_s + F_{ss} \;\;,\label{eq:EBM_surf}\\
c_{ss} \pdiff{T_{ss}}{t} = - F_{ss} \;\;,\label{eq:EBM_subsurf}
\end{align} 
where $a, s, ss$ subscripts denote atmospheric, surface and subsurface
quantities respectively, $c$ is heat capacity (Jm$^{-2}$K$^{-1}$), $S$
is the solar radiation (shortwave), $I$ is net infrared longwave
radiation, $F$ is sensible heat flux (W/m$^{-2}$), and $D_a$ is
meridional heat flux. See table~\ref{Tab:EBM_Parameters} for parameter
values.

We modify this EBM to include radiative forcings from atmospheric
composition and a temperature-dependent precipitation, detailed in
\ref{sec:ebm-modifications}. The atmospheric composition forcings
represent \cotwo{}, water vapour, lapse rate, and cloud effects.
These radiative forcings are treated with two terms: one for the
\cotwo{} forcing, and another for the aggregate effects of water
vapour, lapse rate and cloud forcings.  Both terms are changes in the
mean height at which the atmosphere becomes transparent to longwave
radiation and emits to space, thus adjusting the longwave energy balance.

The net longwave radiation balance of the atmosphere $I_a$ has three terms
representing, respectively, the longwave emissions from the ground
(absorbed by the atmosphere), emissions from the atmosphere to the
ground, and emissions from the atmosphere into space.  Applying the
collective radiative forcings to $I_a$ gives
\begin{align}
I_a = \sigma T^4_{s} - \big(\epsilon_a \sigma  \big(T_a-\Gamma_m
H_{as}\big)^4 +  R_{\textrm{DLW}}\big)
-\:\: \epsilon_a \sigma
\big(T_a+\Gamma_m (H_{ul}+\Delta z_{C} + \Delta z_{\textrm{WLC}})\big)^4 \;\;,\label{eq:I_a_full}
\end{align}
where $\sigma$ is the Stefan-Boltzmann constant, $\epsilon_a$
is atmospheric emissivity, $T_a$ is the temperature of the middle
atmosphere, $\Gamma_m$ is the temperature profile in the atmosphere
$\textrm{d}T/\textrm{d}z$, $H_{as}$ is the middle-atmosphere-to-surface
height, $H_{ul}$ is the default middle-atmosphere-to-upper-layer
height, $\Delta z_{C}$ is the change in upper layer height due to
\cotwo{} concentration in the atmosphere, and $\Delta
z_{\textrm{WLC}}$ the change in upper layer height modelling the
parameterised water vapour, lapse rate, and cloud feebacks.

To validate this reformulation of the \cite{Huybers:2008fg} EBM,
we compare our model against present-day climate, and
perform a \cotwo{}-doubling experiment.

\begin{figure}[ht]
  \centering
  \includegraphics[width=13cm]{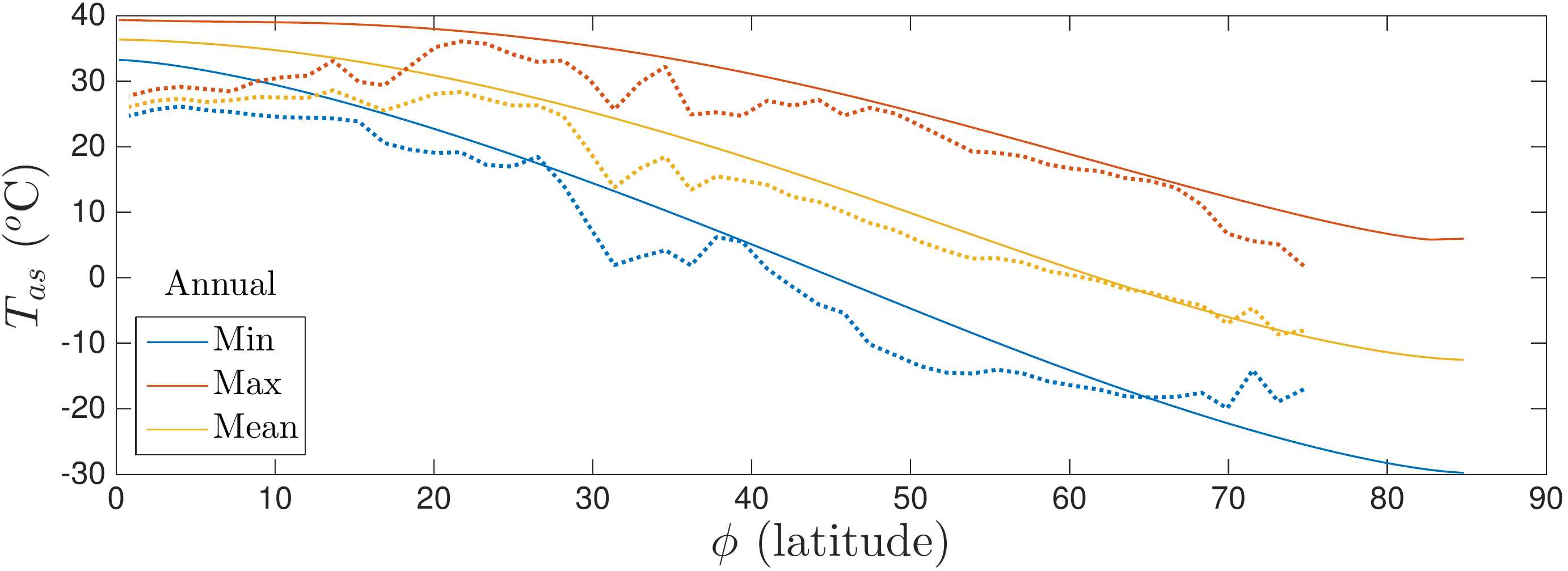}
  \caption{Annual mean, maximum, and minimum surface air temperatures
    for our model (solid lines) and average 1950-80 land data (dashed
    lines) \citep{BerkeleyTemp}. Model uses fixed modern insolation,
    and preindustrial \cotwo{} concentration.  Temperature data is
    based on a transect of land points closest to 52E.  This
    representative transect was chosen as it is low-elevation land,
    removed from oceanic influences, thus replicating our zero relief,
    ocean-free EBM.  Furthermore, this region hosted at ice sheet at
    the LGM.}
  \label{fig:Seasons}
\end{figure}

Figure~\ref{fig:Seasons} shows our model's calculation of
preindustrial conditions and the average surface temperature for
1950-80 \citep{BerkeleyTemp}.  To approximate our model's land-only,
zero-relief Earth, the land temperatures are on a transect of data
points closest to 52E --- a continental regime with minimal ocean
influence and low relief.  Annual mean, maximum, and minimum
temperatures are well aligned between our model and the data.  The
largest errors are near the equator, presumably owing to our model
lacking latent heat transport, Hadley cell circulation, a
representation of the low land fraction near the equator ( \ie{} no
longitudinal heat transport) and because of our no-flux equatorial boundary
condition.  However, the seasonal temperature range is accurately
captured across latitudes (except the polar coast, where oceanic
buffering effects slightly reduce the annual temperature range), and
the mean model temperature is within 1~K of the observed record at the
high latitudes (55-75N) most relevant to ice sheet dynamics.  

\begin{figure}[ht]
  \centering
  \includegraphics[width=13cm]{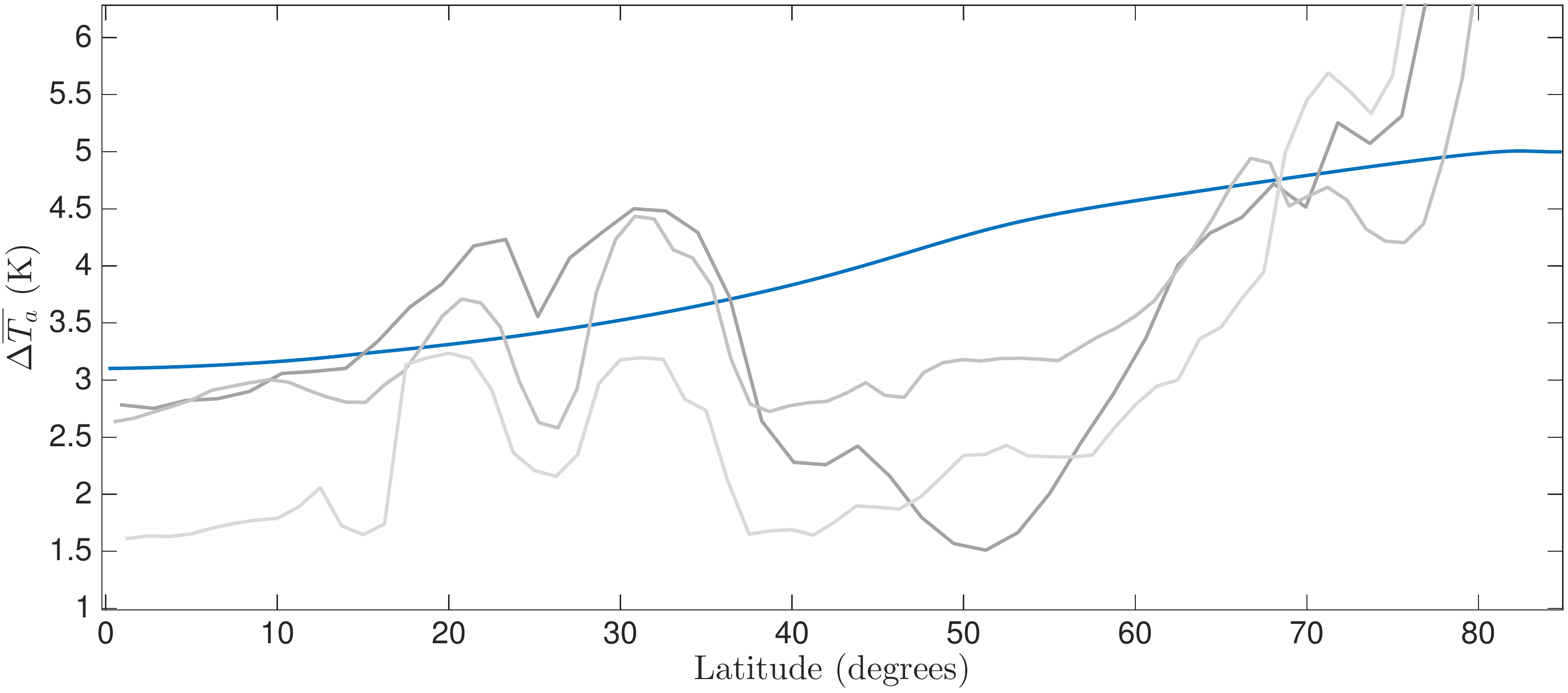}
  \caption{Change in annual average atmosphere temperature for a
    doubling of \cotwo{} from 280 to 560 ppmv under ice-free
    conditions with present-day insolation in our model (blue) and for
    \cotwo{}-doubling experiments in CMIP3 (grey). From dark to light
    grey these are ECHAM, miroc 3.2, and hadgem1; to reduce noise,
    data shown is the difference between the initial and final five
    years of the CMIP models. CMIP data is taken along the same
    transect as in figure~\ref{fig:Seasons}.}
  \label{fig:CO2_Doubling}
\end{figure}

Figure~\ref{fig:CO2_Doubling} shows an experiment in which we double
\cotwo{} from preindustrial conditions, holding atmospheric \cotwo{}
concentration constant and running the model to equilibrium
temperature. We calculate an increase in annual global average
atmospheric temperature of 3.7~K; placing our model within the range
of GCM predictions for \cotwo{}-doubling.  However, we do not match
the some features of the change in temperature with latitude in
GCMs. 1) Temperature anomalies around the ascending/descending arms of
atmospheric circulation cells, and 2) Large polar amplification,
driven by vast reduction in arctic sea-ice
\citep{Rind:1995gf}. However, recent interglacials are not thought
to remove arctic sea ice, so this difference is not important in
modelling late-Pleistocene glacial cycles.

Finally, the EBM includes an ice melting scheme (detailed in
\cite{Huybers:2008fg}) --- whereby if ice-covered ground reaches
$0^{\circ}$C, ice melts according to the available thermal energy ---
giving the annual ice accumulation/melting at each latitude, an input
for our ice sheet model.

\subsection{Ice Sheet Model} \label{sec:ice-sheet-model}
Conceptually, our ice model combines the EBM's annual ice
accumulation/melting with ice flow under gravity and isostatic bedrock
adjustments. We calculate both the evolution of ice thickness across
latitudes $h(t,\phi)$ and the global ice volume $V(t)$. The former is
used by the EBM for ground height and reflectivity, and the latter is
used to calculate volcanic responses to glaciation and sea level.

We use a vertically-integrated 1D model for $h$, following the
\cite{Huybers:2008fg} model exactly, except we use a higher grid
resolution and smaller timesteps.  This ice model calculates the
thickness of a northern hemisphere ice sheet flowing according to
Glen's Law, with an accumulation or ablation of ice at each latitude
calculated according to the EBM.  It assumes incompressible ice,
temperature-independent ice deformability, and the `shallow ice'
approximation whereby deformation is resisted only by horizontal shear
stress, including basal stress.  The ground surface is initially flat,
and deforms to maintain local isostatic equilibrium in each gridcell.
Basal sliding is included via a shearable sediment layer, such that
the base of the ice sheet can move with respect to the bedrock.  Full
details for calculating $h$ can be found in \cite{Huybers:2008fg}, and
for completeness we state the equations and parameter values in
appendix~\ref{sec:ice-sheet-model-appdx}.

   We calculate ice volume from the vertically-integrated 1D ice model by assuming that ice
   sheet width is 60\% of Earth's circumference at each latitude
   $\phi$ --- a reasonable approximation at high northern latitudes. Ice volume is
   expressed in eustatic meters sea-level equivalent (msle) by
   dividing ice sheet volume by the ratio of ice to water density and
   the surface area of the ocean. Over glacial cycles, thermal expansion of the oceans is
   negligible at $<1$\% of glacial sea level change
   \citep{McKay:2011gs} and we ignore it in our sea level calculation.

\subsection{Carbon Model} \label{sec:carbon-model} 
The last component of our model calulates the \cotwo{} concentration
in the atmosphere over time, responding to changes in climate
configuration and volcanic emissions.  We will consider the \cotwo{}
influences in turn, and discuss their timescale and magnitudes.

Glacial--interglacial \cotwo{} variations are not fully understood,
and certainly cannot be replicated from first principles.  Therefore
we circumvent the accounting of all sources and sinks of \cotwo{}. We
instead parameterise atmospheric \cotwo{} concentration, $C$, as
proportional to average global temperature, matching a
well-established feature of reconstructed Pleistocene climate records
\citep{Cuffey:2001vd,Sigman:2010bz}.

This carbon--temperature feedback accounts for all potential feedbacks
in the surface carbon system, such as ocean-atmosphere equilibration and
biosphere changes, and aggregates them to a single feedback
parameter.  This simplification allows us to be agnostic about
the causes of these \cotwo{} changes and to enforce agreement with
the observed correlation between \cotwo{} and ice volume in the
Pleistocene. However, it fails to capture, for example, state
dependency (a Kelvin change in average planetary temperature 
changes atmospheric \cotwo{} by a fixed amount, regardless of the
current temperature).  This may be important given recent
suggestions of a lower limit on $C$ during glacial
cycles \citep{Galbraith:2017vq} and several plausible non-linear
components partitioning \cotwo{} in the surface system. These
include, but are not limited to, hysteresis in the ocean
overturning circulation \citep{Weber:2007va}, iron fertilisation
\citep{Watson:2000ws}, plant growth being non-linearly temperature
dependent, and seafloor and permafrost methane clathrate release
\citep{MacDonald:1990vd}.   Despite these complications, the
overall linear $C,T$ relationship in the Pleistocene suggests our
formulation is a good representation of leading order behaviour.

We also include changes to atmospheric \cotwo{} from volcanic
emissions as separate, independent terms, giving a carbon equation:
\begin{align}
\pdiff{C(t)}{t} = \gamma_{\textrm{T}} \pdiff{\overline{T}_s}{t}  +
\gamma_{\textrm{MOR}} \;  f_{\textrm{MOR}}\left( \pdiff{V}{t} \right)   +
\gamma_{\textrm{SAV}} \;  f_{\textrm{SAV}}\left( \pdiff{V}{t} \right) 
\;\;, \label{eq:Carbon_full}
\end{align}
where $f_{\textrm{SAV}},f_{\textrm{MOR}}$ are functions that map sea
level history to current \cotwo{} emissions for global subaerial and
mid-ocean ridge volcanism respectively. The $\gamma_i$ are
coefficients that represent the sensitivity of $C$ to changes in the
Earth system. The \gT{} term denotes the sensitivity to changes in
surface temperature; this is interpreted physically as the net
effect of surface system (\ie{} atmosphere, biosphere, and ocean)
partitioning of \cotwo{} between the atmosphere and other
reservoirs. \gT{} has units of \cotwo{} mass per Kelvin change in
(annual and spatial) average planetary temperature, stated in ppmv/K
for convenience (7.81 Gt\cotwo{} = 1 ppmv change in atmospheric
\cotwo{}
concentration). %
The \gMOR{} and \gSAV{} coefficients are $C$ sensitivity to changes in
sea level caused by variable MOR and subaerial volcanic \cotwo{}
emissions. These coefficients state the peak change in annual volcanic
\cotwo{} emissions resulting from a given rate of sea level change,
and thus have units of Mtonnes \cotwo{} per year per cm/yr change in
sea level. 
Volcanic \cotwo{} emissions have distinct timescales for subaerial and
MOR volcanic systems. MOR \cotwo{} emissions, according to modelling
by two of the authors \citep{Burley:2015ic}, respond to glacial
sea-level change with a tens-of-thousands-of-years lag.  Subaerial
volcanism responds comparatively fast to changes in nearby ice sheets,
with field evidence \citep{Rawson:2015ug,Kutterolf:2013dd} showing
responses in approximately 4~kyrs.

\begin{figure}[ht]
  \centering
  \includegraphics[width=12cm]{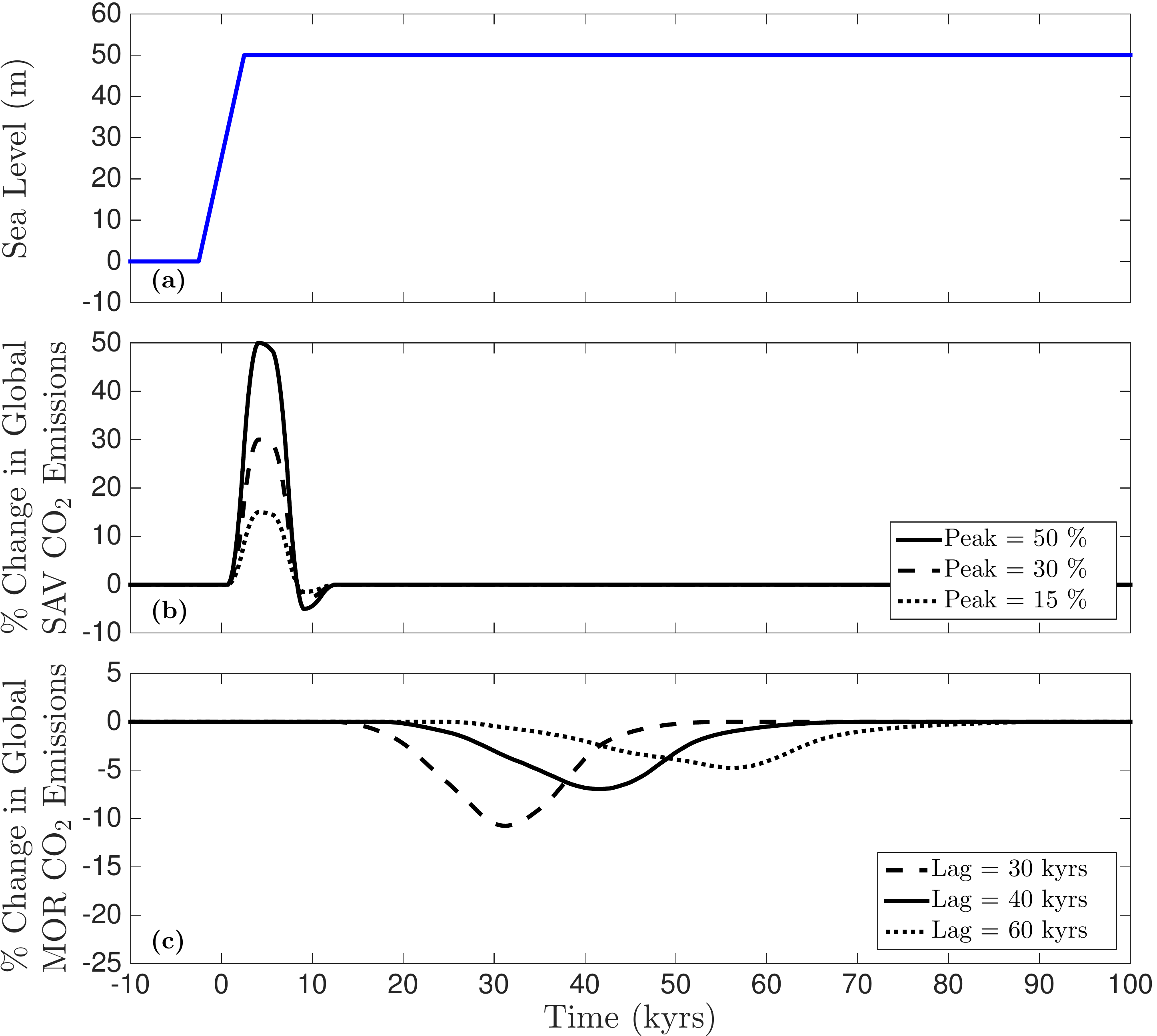}
  \caption{Toy deglaciation event with plots of \textbf{(a)} sea level, and the
    consequent \cotwo{} emissions of \textbf{(b)} subaerial and \textbf{(c)} MOR
    volcanism. $\gMOR$ and $\gSAV$ set the peak values of the
    emissions in panels (b),(c).  The integrals of \cotwo{} emissions
    curves in panel (c) are all equal.}
  \label{fig:Dual_Greens_Example}
\end{figure}

As shown in figure~\ref{fig:Dual_Greens_Example}, we use an approximate
Green's function representation of each system where the rate of
change of global ice volume (directly proportional to sea level)
produces a change in \cotwo{} emissions at a later time. The \gMOR{}
and \gSAV{} coefficients scale the height of these Green's functions.  The
reasoning behind the imposed temporal patterns and magnitude of
volcanic response is explained below.

The MOR Green's function follows the global MOR results in
\cite{Burley:2015ic}. There are no published observational
constraints on MOR \cotwo{} response to glaciation that can support or
reject this model.  However, records of sea-floor bathymetry are
consistent with a sea-level-driven MOR eruption volume model
\citep{Crowley:2015ec} that shares many features with the
\cite{Burley:2015ic} model (though see \cite{Olive:2015eq}). 

In \cite{Burley:2015ic} sea-level change causes a \cotwo{} anomaly in
mantle melt at about $60$~km depth below the MOR.  This \cotwo{} anomaly is
subsequently carried by magma to the MOR axis.  The MOR Green's
function's magnitude and lag time are determined by the mantle
permeability $K_0$, a physical property that controls how quickly
mantle melt percolates through the residual (solid) mantle grains.
The mantle permeabilities assumed in this paper are within the
accepted range \citep{Connolly:2009bw}, and give \cotwo{} travel times in
agreement with the $^{230}$Th disequilibria
in MORB \citep{Jull:2002fk}.  

Figure~\ref{fig:Dual_Greens_Example}c shows example MOR \cotwo{}
emissions responses for a range of mantle permeabilities.  They show
similar features: a decrease in \cotwo{} emissions lasting 10s-of-kyrs
that lags the causative sea-level increase by 10s-of-kyrs.  The total
change in \cotwo{} emissions (\ie{} the integral of
figure~\ref{fig:Dual_Greens_Example}c) is the same for all
permeabilities.

In subsequent sections, we
discuss behaviour in terms of the `MOR lag' rather than mantle
permeability, as the former has a more direct interpretation that
is relatable to other model components (as in
fig~\ref{fig:MORC_didactic}).

MOR \cotwo{} emissions dissolve into intermediate ocean waters,
delaying entry into the atmosphere by a few hundred years. This delay
is much smaller than both the MOR lag time and the
uncertainties therein; hence we neglect it.

The SAV Green's function has a temporal pattern based on the
observation-derived eruption volume calculations in
\cite{Rawson:2015ug,Rawson:2016wa}; these show a large increase in
eruptive volume per unit time (volume flow rate) 3--5~kyrs after
deglaciation, followed by a few kyrs of low eruptive volume per unit
time, then a return to baseline activity. This timing is consistent
with other studies that report an increase of subaerial arc volcanism
that lags behind deglaciation by $\sim$$4$~kyrs
\citep{Jellinek:2004hb,Kutterolf:2013dd}. Therefore, the volume-flow-rate
timeseries of \cite{Rawson:2015ug} represents the temporal response of
SAV accurately. 

However, this is the response of a single volcano, and
we need to model the global volcanic system.  The planet's volcanoes
experience different glacial coverage during an ice age, so the change
in a single volcano's volume-flow-rate is not a valid basis for a global aggregate.
Therefore, we want to adjust the magnitude of volume-flow-rate change,
while keeping the temporal response pattern. 

We create a
representative global value for the volume-flow-rate change by using the
eruption frequency datasets of \cite{Siebert2002} and
\cite{Bryson:2006tn}, as compiled in \cite{Huybers:2011fr}.  To do
this, we assume that eruption frequency is proportional to eruptive
volume per unit time. This is an oversimplification, however eruption
frequency is the only available constraint on global subaerial
volcanic behaviour over a glacial cycle (erosion, reworking, and
burial of volcanic units causes great difficulties in eruption volume
calculations prior to the past few thousand years).  Eruption
frequency increases by at least $\sim$50\% during deglaciation. Next,
we consider how to calculate the SAV \cotwo{} emissions.

To relate SAV eruption volume per unit time to \cotwo{} emissions
there are three regimes to consider: if increased SAV volcanic
eruption volume during deglaciation is entirely due to venting of
pre-existing magma reservoirs, there would be direct proportionality
between \cotwo{} flux and eruption volume; at the other extreme, if
the eruption-volume increase is entirely due to enhanced melting of a
\cotwo{}-depleted mantle there is, to leading order, no correlation
between eruption volume and \cotwo{} flux (see \cite{Burley:2015ic}
appendix~A.4).  Finally, if there is variable melting of a
carbon-bearing phase (either mantle or metamorphism of a crustal rock
unit \citep{Goff:2001br}) there will be a correlation between eruption
volume and \cotwo{}, but of unknown strength and with a dependence on
location. For lack of information to guide us, we model SAV \cotwo{}
emissions as directly correlated to the rate-of-change of ice
volume. This leaves considerable uncertainty in the coefficient
$\gSAV$.

Finally in volcanism, it is unclear if hotspots have a
glacially-driven variability.  Their deep melting systems
\citep{hardhardottir2017spatial,Yuan:2017gb,Zhao:2001um} preclude
sea-level and glaciation influencing depth-of-melt-segregation as at
MORs.  Instead, hotspots' extensive magma chamber systems
\citep{hardhardottir2017spatial,Larsen:2001ww} imply they respond like
SAV.  This is consistent with the observed volume flow rate in Iceland
\citep{Maclennan:2002ba} (although other mechanisms could also be
consistent with the data).  We expect hotspots respond on the same
timescale as arc volcanoes; however, most hotspots are oceanic thus
the pressure change will be caused by sea level rather than ice
sheets. Thus hotspots (if they have any glacially-driven \cotwo{}
variability) will be a negative feedback acting simultaneously with
arc volcanism, thus increasing the uncertainty in the appropriate
value of \gSAV{}.

Above, we have described the logic leading to our Green's function
representations of MOR and subaerial volcanism.  The physics-driven
and data-driven calculations in this logic prescribe the percentage
change in \cotwo{} emissions in response to rate-of-sea-level-change.
We multiply the percentage value by the average annual volcanic
emissions to get Green's
functions in units of Mt\cotwo{}/year per cm/yr.   Therefore, the
Green's functions' magnitudes have uncertainty from both the
calculated percentage change and the default emissions value.

Annual MOR \cotwo{} emissions have large uncertainties, with papers
stating 2-standard-deviation lower bounds of 15--46 Mt\cotwo{}/yr, and
upper bounds of 88--338~Mt\cotwo{}/yr from geochemical analyses
\citep{Marty:1998vo,Resing:2004ks,Cartigny:2008dz,Dasgupta:2010fw}.
The most recent estimates by \cite{LeVoyer:2017cb} are MOR \cotwo{}
emissions of 18--141~Mt\cotwo{}/yr.  The $91$~Mt\cotwo{}/yr estimate
used in \cite{Burley:2015ic} is fairly central in that range and for
consistency I continue to use that value in this thesis.

Annual SAV \cotwo{} emissions are also uncertain.  Studies estimate
SAV \cotwo{} emissions are within $\pm15$\% of MOR \cotwo{}
emissions \citep{Marty:1998vo,FISCHER:2008dq}, much less than the
uncertainty in each value.   For simplicity, we set background SAV \cotwo{}
emissions as equal to MOR \cotwo{} emissions.

We assume that the solid Earth has no net effect on atmospheric
concentration of \cotwo{}, $C$, over the late
Pleistocene, and therefore when SAV or MOR volcanism are at baseline
emissions (\ie{} $0\%$ in figure~\ref{fig:Dual_Greens_Example}(b,c))
they do not affect $C$.  Any increase or decrease from average
volcanic \cotwo{} emissions acts to increase or decrease $C$.
Physically, this assumes the weathering drawdown of
\cotwo{} balances the time-average of volcanic emissions, and
that any variations in the weathering rate are at the sub-ka
timescale (captured by \gT{}) or negligible on the 1~Ma timescale.

The volcanic Green's functions assume that all volcanic \cotwo{}
variations directly change \cotwo{} concentration in the atmosphere.
However, we might expect, for example, an extra 5~Mt/yr of volcanic
\cotwo{} to be partially absorbed by the ocean such that atmospheric
\cotwo{} mass does not increase at 5~Mt/yr.  For modern oceans it is
calculated that $15$--$30\%$ of \cotwo{} added to the atmosphere
remains after 2~kyrs \citep{Archer:2009bm}. However, such calculations
are state dependent; both the decay timescale and equilibrium airborne
fraction vary with the injected \cotwo{} mass and the initial ocean
state. There are no estimates of the decay timescale or equilibrium
airborne fraction on glacial timescales, nor glacial--interglacial
ocean models from which one could be extracted.  For simplicity, plots
in this paper assume that all volcanic \cotwo{} remains in the
atmosphere, however it is perhaps fairer to discount emissions by a
constant factor --- this discount is discussed in the conlusions
section in terms of the of volcanic emissions required for certain
climate behaviour.

Finally, we highlight a feature of the volcanic response that is
important for understanding $C$ evolution over time in
equation~\eqref{eq:Carbon_full}: the total change in MOR \cotwo{}
emitted (the integral of curves in
figure~\ref{fig:Dual_Greens_Example}c) is directly proportional to the
amplitude of sea level change \citep{Burley:2015ic}.

 Therefore the amplitude of changes in atmospheric \cotwo{}
 concentration $\Delta C$ caused by volcanism, for a single change in
 sea-level, are directly proportional to the amplitude of sea-level
 changes $\Delta V$ (section~\ref{sec:mor-co_2-response} illustrates
 the more complex scenario of periodic sea-level).  By comparison, $C$
 changes due to surface system feedbacks are proportional to changes
 in mean atmospheric temperature $\bar{T}$. $C$ determines radiative
 \cotwo{} forcing and thus this forcing depends upon past variations
 in ice volume $V$ and $\bar{T}$.  Furthermore, the effective
 insolation forcing depends on planetary albedo (\ie{} ice sheet
 extent).  Consequently, the balance of climate forcings in the model
 varies as the amplitude of changes in ice sheet volume, extent, and
 mean atmospheric temperature vary.

 Note that we only model a single variable volcanic process ---
 \cotwo{} emissions --- yet other glacially-driven volcanic changes
 could affect climate.  For instance: 1) SAV aerosol emissions will
 increase following deglaciation.  This could be a positive or
 negative climate feedback depending on injection height, particle
 size and composition distributions --- all poorly constrained even
 for current volcanic systems.  2) MORs will have varying emissions of
 many chemical species in response to glacial cycles, \cotwo{} only
 represents the end-member of highly incompatible species (partitions
 strongly into mantle melt), with less incompatible species having a
 shorter lag and smaller variation than \cotwo{}.  Some species, like
 bio-active Fe or those that may affect ocean pH, could be relevant
 to global climate.  3) MOR hydrothermal systems vary with glacial
 cycles \citep{Lund:2011jd, Middleton:2016fz}. Increased melt
 productivity at MORs would presumably drive more vigorous convection
 of seawater in the hydrothermal system. However, the net effect of
 increased circulation is hard to predict due to the complex chemical
 and biological processes acting on fluid composition.

 These potential glacially-driven volcanic effects have large
 uncertainty and complex underlying physical processes.  We choose to
 not include them; they would increase model complexity and lead to an
 excess of uncertain parameters with overlapping timescales.

Having defined the component models, we now describe the coupling
between these components and how the combined model is initialised.

\subsection{Coupling and Initialising the Model}\label{sec:coupl-init-model}
The three component models operate on different timescales and hence
it is not immediately clear how to best couple them together. Careful
consideration of timescales will inform our choice.

The fastest changes in the model are the seasonal changes in
insolation and temperature, setting the shortest timestep in the model
at $10^{-2}$~years. Taking such small timesteps for a full million
years would be prohibitively expensive, so we use the simplification 
that $1)$ annual averages of thermal quantities are accurate drivers
of ice sheet flow and carbon change (for example, we calculate ice
sheet growth using the annual average melting/accummulation rate), and
$2)$ %
subsequent years are very similar.  Consequently we use the EBM model
to calculate the equilibrium temperature and precipitation/melting
distribution for the current \cotwo{} concentration and ice sheet
configuration.  We then hold temperature and precipitation constant while
running the carbon and ice sheet models.  After small changes in $C$
and ice configuration we run the EBM again, calculating a new temperature
 and precipitation/melting distribution to drive further changes in $C$ and ice.

  The timescale for these `small changes' in ice and $C$ will clearly
  be greater than a year.  In testing the model, we found a timescale
  on the order of $200$~years was suitable. Shorter timescales do not
  alter $V$ or $C$ significantly.

  For the results presented here, the ice model was run for
  intervals of 250~years, with two-year timesteps.  The carbon
  concentration in the atmosphere is updated every 250 years, then the
  EBM is run for five years to update the temperature and
  precipitation/melting distribution in preparation for continuing the
  ice model. 
  To initialise the model at a particular time in the past, the
  insolation is computed for that time. The \cotwo{} concentration is
  taken from ice core data.  These are both held constant while the
  EBM and ice sheet come into equilbrium.  Subsequently, the model is advanced
  using the timestepping described above.

The range of fully-defined initialisation times are limited by the
atmospheric \cotwo{} record, which extends back $800$~kyrs
\citep{Bereiter:2015js} (insolation is well defined for 10s-of-Myrs
\citep{Berger:1991fm,Laskar:2004fc}).

We could use earlier start times by solving an inverse
problem to define starting $C$: use a range of initial $C$ values and
match the resulting equilibrium ice sheet volume to a proxy sea-level
record. However the difficulties and objections such a method
raises would distract from the core investigation of this paper.

\clearpage{}
\section{Results: Basic Model Behaviour}\label{sec:results}
\subsection{Mid-ocean ridge CO$_2$ response to sinusoidal sea level}\label{sec:mor-co_2-response}
This section demonstrates how global MOR CO$_2$ emissions respond to
sinusoidal sea-level changes, neglecting climate feedbacks
from that \cotwo{} change. 

\begin{figure}[ht]
  \centering
  \includegraphics[width=12cm]{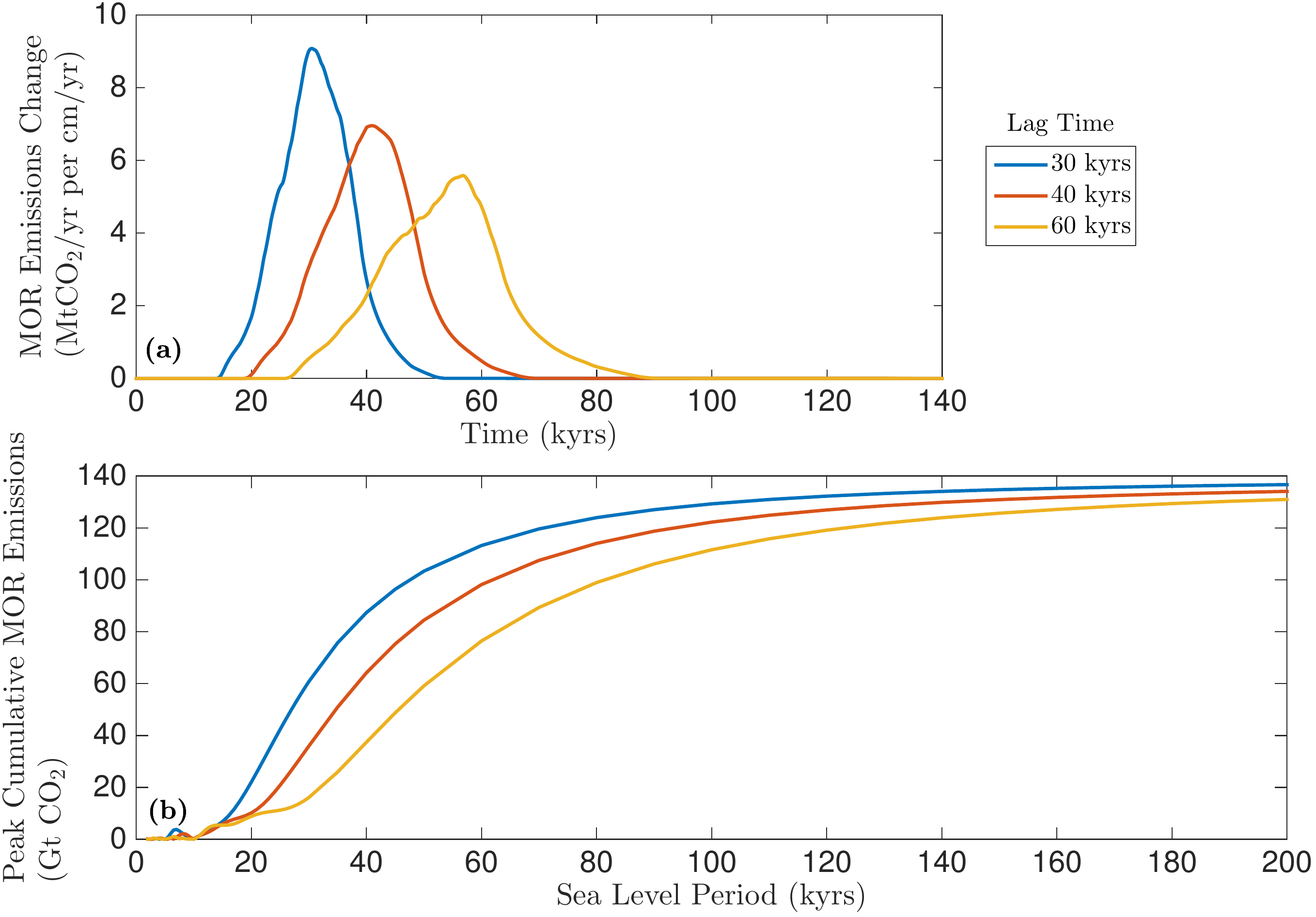} 
  \caption{MOR CO$_2$ emissions driven by sinusoidal sea level of 50~m
    amplitude, no climate feedbacks.  Panel \textbf{(a)} shows Green's
    functions for the global MOR system with different lag times;
    stated as the global MOR emissions change in MtCO$_2$/yr per cm/yr
    rate-of-change in sea level. The \gMOR{} values are those
    predicted from \cite{Burley:2015ic} for the lag times shown: 9.2,
    7.1, 5.6~Mt\cotwo{}/yr per cm/yr rate-of-change of sea-level.
    Panel \textbf{(b)} shows the maximum cumulative MOR CO$_2$
    emissions.  7.81 Gt\cotwo{} = 1 ppmv $C$ change, so maximum values
    in panel (b) are equivalent to 18 ppmv.}
  \label{fig:MORC_periodicSL}
\end{figure}

As shown in figure~\ref{fig:MORC_didactic}, sinusoidal sea-level
causes a sinusoidal variability in relative MOR CO$_2$ emissions rate
(Mt \cotwo{} per year relative to baseline MOR emissions).  When these
relative \cotwo{} emissions are positive, MORs are increasing the
\cotwo{} concentration in the atmosphere; when negative, \cotwo{}
concentration in the atmosphere is decreasing.  Therefore, taking the
integral (with respect to $t$) of the relative MOR \cotwo{} emissions
rate gives the total change in atmospheric \cotwo{} mass caused by
MORs, which is also sinusoidal.  The peak-to-trough magnitude of this
sinusoid (after a transient windup period) is the `maximum cumulative MOR
CO$_2$ emissions' --- the maximum \cotwo{} mass that variable \cotwo{}
emissions add the atmosphere.
Figure~\ref{fig:MORC_periodicSL}b shows maximum cumulative MOR CO$_2$
emissions across a range of sinusoidal sea-level periods, for the
three MOR lag times shown in figure~\ref{fig:MORC_periodicSL}a.  The
maximum cumulative MOR CO$_2$ emissions vary with
sinusoidal sea-level frequency, meaning that MOR \cotwo{} emissions
can have significantly larger effects on $C$ if the period of sea
level change increases.
The physical reason for this behaviour is that the mantle melt (and
associated \cotwo{} anomaly) arriving at the MOR at any given time is
an amalgamation of mantle melts generated at the base of the melting
region across a range of times (the width of the Green's functions in
figure~\ref{fig:MORC_periodicSL}a) in the past. If this range of times
is greater than the sinusoidal sea-level period then \cotwo{}
anomalies of opposing effect arrive at the MOR simultaneously,
reducing variability in MOR \cotwo{} emissions (see
\cite{Burley:2015ic}).  Therefore, as shown in
figure~\ref{fig:MORC_periodicSL}, sinusoidal sea-level periods shorter
than the Green's function width cause small amplitude cumulative MOR
CO$_2$ emissions.

For sinusoidal sea-level periods much larger than the Green's function
width, the cumulative MOR CO$_2$ emissions reach a constant value.
When sea level falls,
the depth of first melting under the MOR increases, creating new melt in a
 deeper section of mantle, and extracting the carbon from that
mantle. The change in depth of first melting (and thus the volume of
mantle decarbonated) is proportional to the amplitude of sea-level
change.  Therefore the maximum possible \cotwo{} injected into the
atmosphere is determined by the amplitude of sea-level change, and
different period sinusoidal-sea-levels have different effectiveness at
reaching this maximum.  Sea-level periods much longer than the Green's
function width allow this mantle volume to degas its carbon and emit
\cotwo{} from the MOR without interference from opposing \cotwo{}
anomalies. Thus
longer sea-level periods converge to the maximum possible \cotwo{}
release into the atmosphere.

Our arguments above state that maximum cumulative MOR \cotwo{} emissions
will have near-zero values for sea-level periods much less than the
width of the MOR Green's function, and converge to a large value for
sea-level periods greater than the width of the MOR Green's
function.
The widths of our Green's functions in
figure~\ref{fig:MORC_periodicSL}a are approximately $30$--$60$~kyrs and,
consequently, figure~\ref{fig:MORC_periodicSL}b demonstrates
significant changes in the amplitude of cumulative emissions over
glacial-cycle-relevant changes in sea-level period.  For example, MOR
systems with 40~kyr lag driven by 40, 80, 120~kyr sea-level period
have maximum cumulative emissions of 64, 104, 126~Gt \cotwo{},
corresponding to a doubling of MOR-derived \cotwo{} deviations when sea
level changes from early-Pleistocene to late-Pleistocene periodicity.

This result is robust for any MOR system with a lag likely to
destabilise 40 kyr glacial cycles (30--50~kyr): 1.4--$2.5\times$
increases in maximum cumulative MOR CO$_2$ emissions if sea level
periodicity increases to $\sim$100~kyrs.  See
appendix~\ref{sec:convolution} for generalised mathematical treatment.
Whilst MOR \cotwo{} emissions remain a small part of the overall
glacial \cotwo{} cycles, this is a mechanism for MOR volcanism to
reinforce $\sim$100~kyr glacial cycles if they occur.

Our SAV Green's function width is 3.5~kyrs, much less than glacial
sea-level periods, thus our calculated cumulative SAV \cotwo{}
emissions do not vary significantly with sea-level period.

\subsection{Forcing with historical CO$_2$ values} \label{sec:IceCoreRep}
In this section we consider a forcing based on reconstructed
atmospheric \cotwo{} concentration, $C$, and insolation. For this
scenario $C$ is set to ice core values, rather than evolving according
to equation~\eqref{eq:Carbon_full}.  The scenario has two purposes: \textit{1)}
validating our EBM and ice sheet model --- the calculated ice sheet
volume $V$ should approximate reconstructed sea-level data, and \textit{2)}
demonstrating our model's $V$ response to $\sim$$100$~kyr $C$ cycles
--- a benchmark for $V$ when subsequent sections calculate $C$
according to equation~\eqref{eq:Carbon_full}.  These are both discussed
below.

\begin{figure}[ht]
  \centering
  \includegraphics[width=15cm]{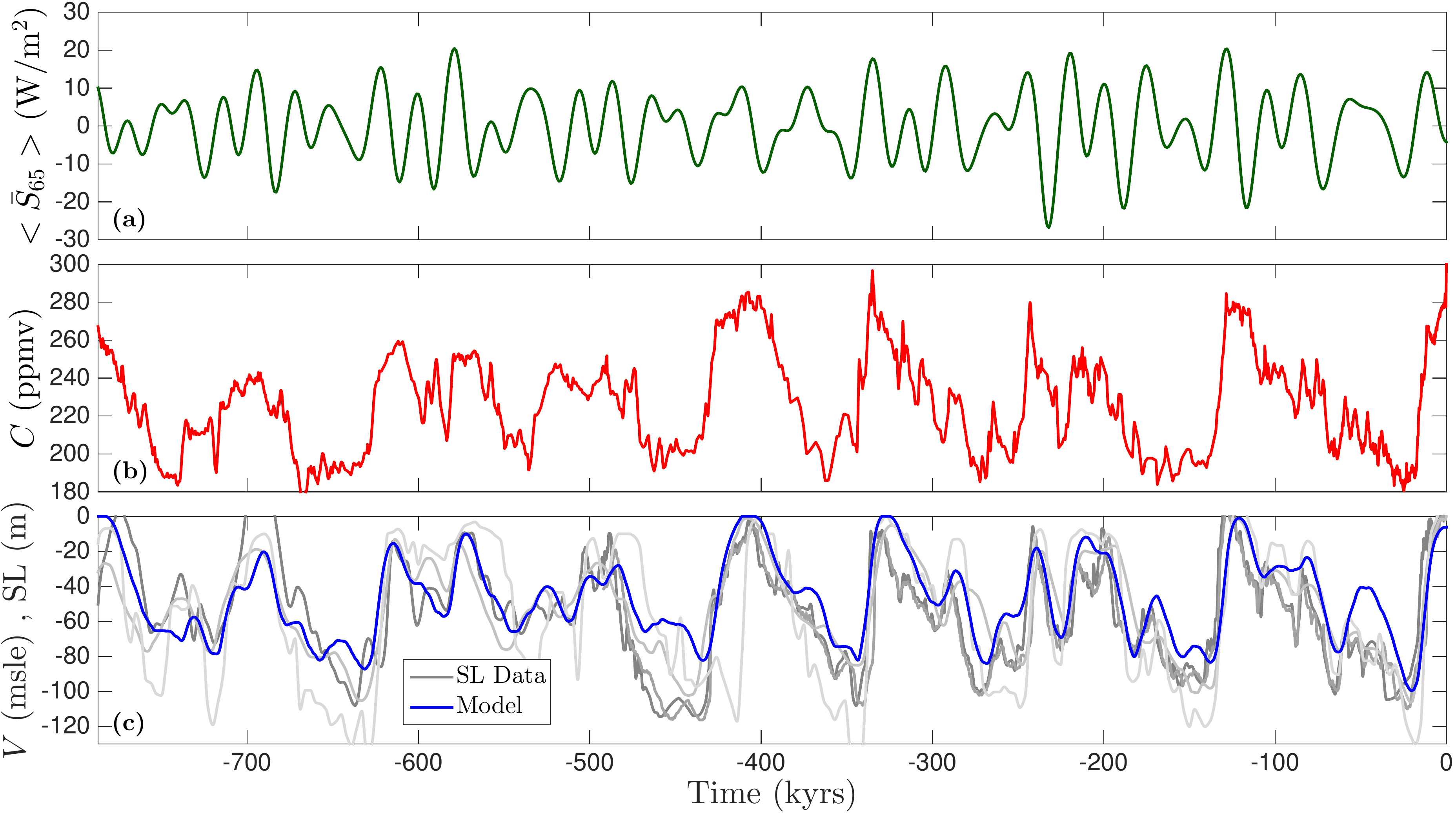}
  \caption{ Model driven by insolation and ice core \cotwo{} values.
    Panel \textbf{(a)} shows mean summer half-year insolation at 65$^{\circ}$N
    as deviation from the mean \citep{Berger:1991fm}. Panel \textbf{(b)} shows
    atmospheric \cotwo{} concentration \citep{Bereiter:2015js}. Panel
    \textbf{(c)} shows model ice volume $V$ in metres sea level equivalent and
    reconstructed sea level from several sources.  From dark to light
    grey these are \cite{ Rohling:2009ez, Elderfield:2012cf} as
    compiled in \cite{MartinezBoti:2015kb}, maximum probability
    Red-Sea record from \cite{Grant:2014sea}, and the \dOxy{} record
    decomposed into a sea level component by \cite{vandeWal:2011kc}
    and \cite{siddall10}.  Discrete Fourier transform of $V$ gives
    power of 49.5\% at 80--120~kyr periodicity, 23.4\% at 40~kyr
    periodicity and 5.6\% at 23~kyr periodicity.}
  \label{fig:IceCoreRep}
\end{figure}

Figure~\ref{fig:IceCoreRep}a and \ref{fig:IceCoreRep}b show the
insolation and $C$ timeseries, and figure~\ref{fig:IceCoreRep}c shows
the calculated ice volume $V$ (blue).  Ice volume is correlated to
both insolation and $C$, as expected.  Furthermore, the calculated ice
volume is a good fit to reconstructed sea-level records (grey).  The
model's most significant difference from sea-level records is a lower
variability at high frequencies; part of this difference is noise in
the data but part is probably rapid changes in ice sheets that our
model does not capture.  Despite this, overall the timeseries in
figure~\ref{fig:IceCoreRep}c are similar.

This similarity suggests the radiative forcings our model adds to
\cite{Huybers:2008fg} are reasonable; we calculate realistic ice sheet
configurations for actual insolation and $C$ values. There are
uncertainties in our WLC radiative forcing parameter (water vapour,
lapse rate and clouds --- discussed in
section~\ref{sec:adding-WLC-forcing}) due to the range in the tuning
GCM cohort's climate sensitivities. Across the plausible WLC forcing
range, the maximum glacial varies from 75~m to 104~m. Changing WLC
forcing does not introduce any novel model behaviour nor change the
timing of turning points in $V$. Therefore we have confidence that our
model behaviour is not contingent on peculiarly specific values of WLC
forcing, and that our chosen value is physically plausible.

Figure~\ref{fig:IceCoreRep}c is a diagnostic for real-world glacial
cycles in our model, demonstrating the ice volume timeseries that
results from late-Pleistocene $C$ and insolation. Our model calculates
powers in the ice volume timeseries, at the 23, 41, and $\sim$100~kyr periods of 5.6\%, 23.4\%, and
49.5\% respectively, similar to the average of the displayed sea level data
(4\%, 10\%, 55\%).
In section~\ref{sec:full-model-behaviour}, we compare Fourier
transforms of the ice volume timeseries from
figure~\ref{fig:IceCoreRep}c and our full model system (forced purely
by insolation, with $C$ determined by equation~\eqref{eq:Carbon_full}).

\subsection{Forcing with Individual Historical Values} \label{sec:IceCoreRep_ish}
Another useful benchmark of model behaviour is forcing with just the
insolation (constant \cotwo{}, 240~ppmv), and just the reconstructed atmospheric
\cotwo{} concentration (constant insolation, 800~kyr average), and
comparing to the result of section~\ref{sec:IceCoreRep}

\begin{figure}[ht]
  \centering
  \includegraphics[width=15cm]{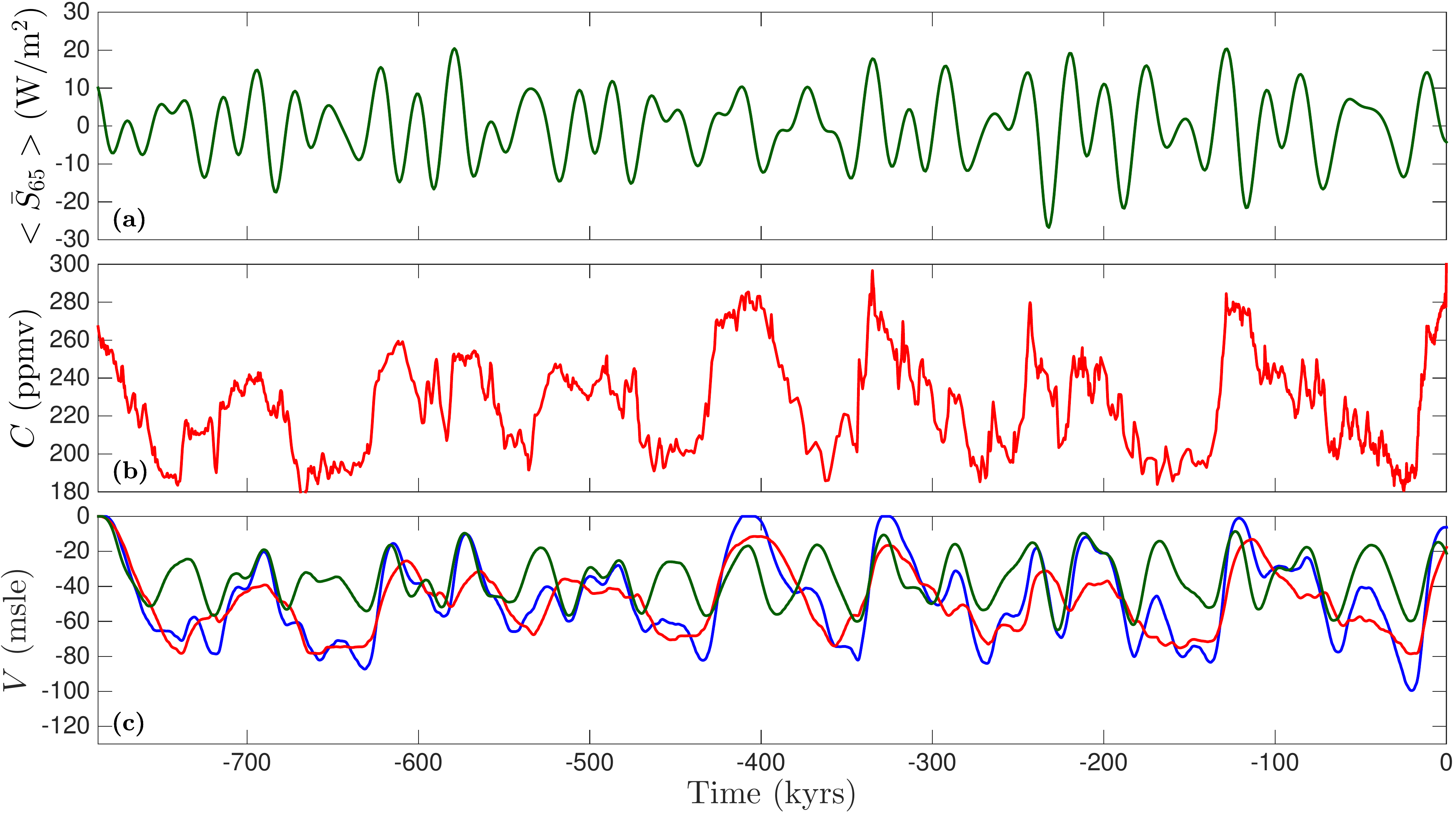}
  \caption{ Model driven by insolation and ice core \cotwo{} values.
    Panels \textbf{(a)}, \textbf{(b)} are identical to
    figure~\ref{fig:IceCoreRep} --- half-year insolation at 65N and
    atmospheric \cotwo{} concentration. Panel
    \textbf{(c)} shows model ice volume $V$ in metres sea level
    equivalent for both forcings (blue), \cotwo{}-only (red) and
    Insolation-only (green).}
  \label{fig:IceCoreRep_ish}
\end{figure}

In figure~\ref{fig:IceCoreRep_ish}, our model calculates that \cotwo{} is the
predominant influence on ice volume for the past 800~kyr (to be clear,
this is not a statement of causality; an imposed \cotwo{} timeseries does
not address the reason for that \cotwo{} variation). Table~\ref{tab:IceCoreRep_Diags}
quantifies this, showing that the power spectrum under both forcings is
a weighted sum of, roughly, 70\% \cotwo{}-only and 30\%
insolation-only spectra.

This is a somewhat surprising result, as the calculations of W/m$^2$
forcing are significantly larger for insolation than \cotwo{} at the
canonical 65N latitude ($\pm$15~W/m$^2$ for mean half-year insolation,
$\pm$4~W/m$^2$ for \cotwo).  However, this is not an apples-to-apples
comparison as \cotwo{} forcing is positive for the whole year, and
insolation is highly seasonal.  Furthermore, insolation at the top of
the atmosphere is not the energy retained by the Earth system, which
in our model is 43.7\% of the top-of-atmosphere forcing for
ice-covered ground (38.2\% for non-icy ground, see
section~\ref{sec:ebm-model}), reducing the effective insolation
forcing on the Earth system to 6.5~W/m$^2$, still $60$\% larger than
the \cotwo{} forcing.

This dominant \cotwo{} effect suggests
an emergent property in the model whereby the year-round \cotwo{}
forcing has a much larger effect on ice-sheets than mean half-year
insolation of a similar magnitude.  This could be due to a magnifying effect
from year-round coherent forcing, or it could be that the 65N
metric does not accurately reflect the forcing on ice
sheets.

\begin{table}[h!]
\begin{center}
    \begin{tabular}{|r|rrr|c|}
\cline{1-5}
  \multirow{2}{*}{Drivers}   & \multicolumn{3}{|c|}{Power} &\multirow{2}{*}{Variance} \\
\cline{2-4}
   & 23~ka & 40~ka & 100~ka &   \\
      \hline
      Inso-only       &	14\% & 70\%  &  3\%  & 190  \\
      \cotwo{}-only & 1\% & 4\%  &  67\%  & 340  \\
      Both	             & 6\% & 23\%  &  50\%  & 550  \\
      \hline
    \end{tabular}
  \end{center}
  \caption{Diagnostic values for ice volume timeseries in
    figure~\ref{fig:IceCoreRep_ish}}
\label{tab:IceCoreRep_Diags}
\end{table}

\section{Results: Volcanic Interactions}\label{sec:volc-interact}

Having explored the basic behaviour of the model, we now show the
effects of including volcanism and a dynamically varying \cotwo{}
concentration to the model.

\subsection{Varying mid-ocean ridge lag} \label{sec:varying-mor-lag}
In this section we determine which MOR lag times disrupt the $40$~kyr
glacial cycles in the model, under simplified pure obliquity
insolation forcing.  This is a quantitative test of
figure~\ref{fig:MORC_didactic}'s hypothesis that 30--50~kyr lags are
capable of disrupting 40~kyr glacial cycles.

For this section, insolation is set to a 41~kyr sinusoidal obliquity
variation, with eccentricity and precession fixed at their
average values over the last 500~kyrs.  Atmospheric
\cotwo{} concentration is only affected by MOR \cotwo{} emissions;
 the temperature and subaerial-volcanism terms in
eqn~\eqref{eq:Carbon_full} are set to zero. However, the cumulative MOR
\cotwo{} emissions change $C$ by about 9~ppmv for 100~m sinusoidal
sea-level at 41~kyr (see figure~\ref{fig:MORC_periodicSL}b), far less than the
100~ppmv glacial--interglacial $C$ change.
Therefore MOR sensitivity to sea-level \gMOR{} is increased to
$10$$\times{}$ the values predicted in \cite{Burley:2015ic},
facilitating $C$ change up to about 90 ppmv.

\begin{figure}[ht]
  \centering
\includegraphics[width=15cm]{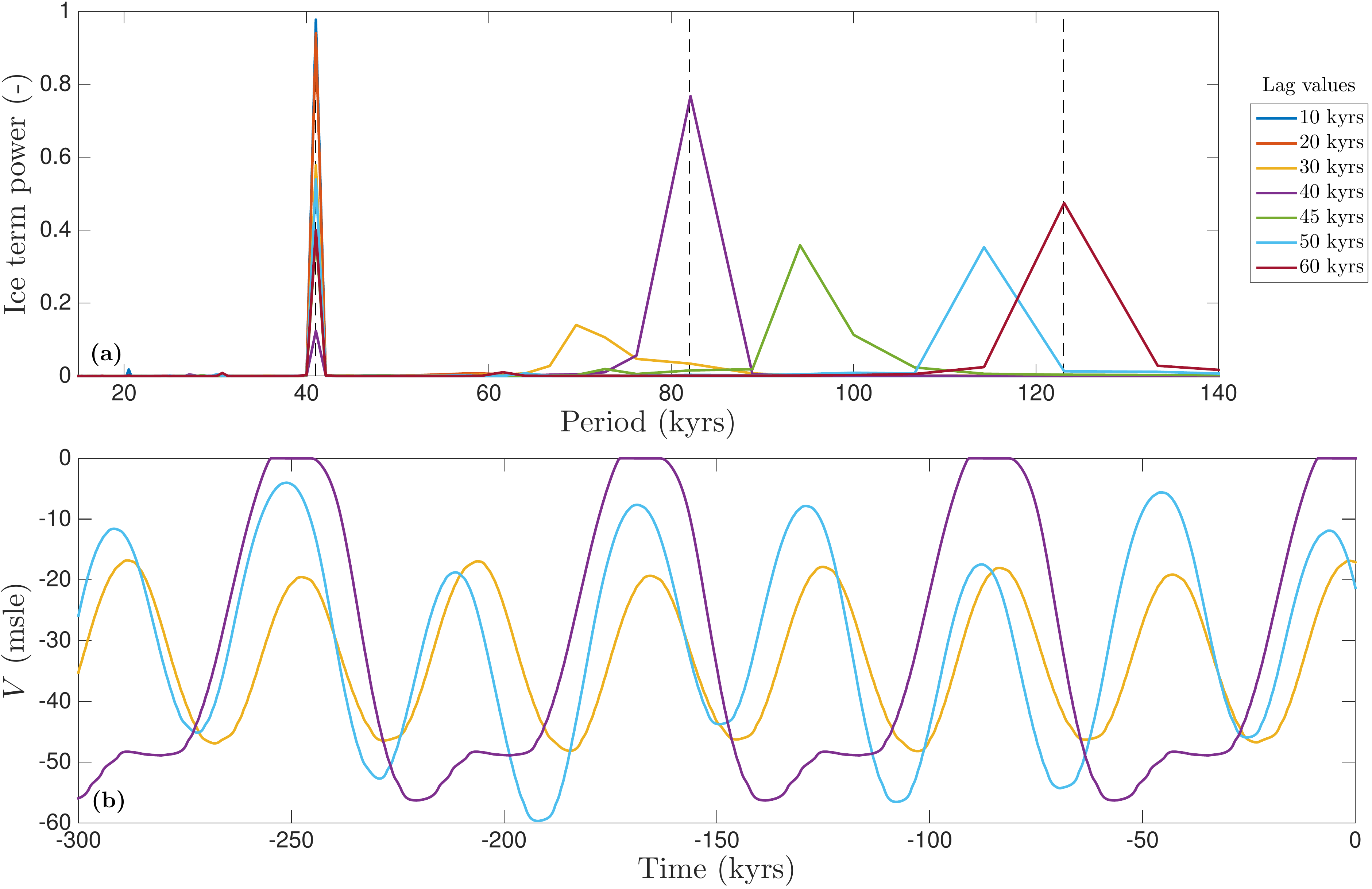}
\caption{ Model for different MOR lag times. Results use
  obliquity-only insolation, with $\gMOR$ non-zero and $\gT , \gSAV
  =0$.  Panel \textbf{(a)} shows the strength of DFT terms for the ice volume
  timeseries. The power spectra terms sum to unity. Panel \textbf{(b)} shows
  select ice volume timeseries over the final 300~kyrs of the model
  run.  The dotted lines mark the 41~kyr obliquity period and its
  multiples.  The mantle permeability range represented by the MOR lag values is
  $10^{-11}$--$10^{-8.5}$~m$^2$ at 1\% porosity. }
  \label{fig:varyingLag}
\end{figure}

Figure~\ref{fig:varyingLag}a shows the changes in ice volume
periodicity for different MOR lag times over a $1.64$~Myr model run,
these results are presented as the power spectrum of ice volume (\ie{}
Discrete Fourier Transform `DFT' of $V$). Figure~\ref{fig:varyingLag}b
shows the final 300~kyrs of $V$ for a subset of these results.  For
lag times less than $20$~kyrs the $V$ cycle remains at 40~kyrs,
phase-locked to the insolation forcing.  Increasing lags across
$20$--$35$~kyr gives the $V$ DFT a subsidiary peak at progressively
longer periods. For 40~kyr lag time the $V$ cycle transitions to
$80$~kyr cycles (80~kyr term is six times the power of $40$~kyr term).
This transition occurs because there are low points in $C$
counteracting every second obliquity-driven deglaciation attempt,
giving the ice volume timeseries shown in
figure~\ref{fig:varyingLag}b.
For lag times $>$$40$~kyr, $V$ cycles have about equal power between
40~kyr and a $>$$80$~kyr cycle.   For a $60$~kyr MOR lag, there is a
dominant cycle at $120$~kyr.

We highlight two features of these results:
firstly, they show that lag times $<20$~kyr do not influence the
periodicity of glacial cycles. This implies that $C$ feedbacks operating at
less than the $<20$~kyr timescale (hereafter, short-timescale) do not
affect the periodicity of glacial cycles.

We term these short-timescale feedbacks because they are shorter than
the obliquity period (\ie{} the default glacial cycle period).  This is an important distinction,
short-timescale feedbacks act on an intra-cycle basis, modulating the
magnitude of glacial cycles and --- in tandem with insolation --- 
controlling the timing of peak climates (see the offset of different
coloured sine peaks in figure~\ref{fig:varyingLag}b).  However, the
short-timescale feedbacks carry little information from one glacial
cycle to the next and are therefore ineffective at disrupting
obliquity-linked 40~kyr cycles.

Thus, short-timescale $C$ feedbacks only affect the magnitude of ice
and temperature changes during glacial cycles; this is true for both
negative $C$ feedbacks (\ie{} acts to oppose sea-level change) shown
in this section and positive $C$ feedbacks (see
figures~\ref{fig:g1_ONLY} and \ref{fig:g3_ONLY}).  Consequently, the
model's glacial cycles are sensitive to the net $C$ change caused by
short-timescale $C$ feedbacks, but relatively insensitive to how the
$C$ change is distributed on very short timescales. This helps justify
our lumping surface system carbon feedbacks into a single parameter,
and suggests we can be agnostic about how carbon feedbacks are
distributed over short timescales if the net carbon change is correct
(\ie{} our model can have inaccurate $\gT$ and $\gSAV$, provided that
their collective effect on $C$ is accurate). This reduces concerns
about the uncertainty of the amplitude of SAV's \cotwo{} response to
glacial cycles.

Secondly, these results largely support the concept that 30--50~kyr
lags can disrupt 40~kyr glacial cycles. A smaller range of
$\simeq$$40$~kyr lag times generate sustained glacial cycles with
80~kyr periods and $\simeq$$60$~kyr lag times generate glacial cycles
with 120~kyr periods.  Of these lag times, the $40$~kyr MOR lag causes
the most power in the $\sim$100~kyr period band, and thus is the optimal
lag for introducing $\sim$100~kyr glacial cycles into an
obliquity-dominated Earth system. To streamline results and
discussion, we use this optimal $40$~kyr lag time in subsequent
sections.  However, under real orbital forcing with power across a
range of frequencies we expect a small range of lags to be similarly
effective at disrupting 40~kyr cycles, because exact (anti-)resonance with
40~kyr orbital cycles will be a relatively less important effect.

These results provide us with the optimal MOR lag time for creating
$\sim$100~kyr glacial cycles, and demonstrate that our model system
has no inherent 100~kyr periodicity until MOR \cotwo{} responses are
introduced as an intercycle feedback.
With the MOR lag time chosen, we now consider the effects of varying
the strength of terms in our \cotwo{} feedback
equation~\eqref{eq:Carbon_full}.

\clearpage{}

\subsection{Full model behaviour} \label{sec:full-model-behaviour}
In this section all terms in our \cotwo{} model
(eqn~\eqref{eq:Carbon_full}) are active; atmospheric \cotwo{}
concentration, $C$, varies according to our parameterised surface
system and volcanic effects. Insolation forcing includes obliquity,
precession, and eccentricity.  We refer to this as the `full model'
configuration. We explore model behaviour by varying all three
sensitivity parameters in the \cotwo{} equation
(appendix~\ref{sec:tuning-c-feedback} shows the model with only single
carbon terms active). We first observe behaviour with timeseries
plots, then use Fourier transforms of ice volume, $V$, to highlight
the changing periodicity of ice volume as $C$ feedback parameters are
changed.  Our benchmark for $\sim$100~kyr cycles is
figure~\ref{fig:IceCoreRep}c --- the $V$ timeseries calculated by our
model when forced by both an imposed, ice-core \cotwo{} timeseries and
insolation. We compare our DFT terms from the full model to this `ice
core replication' benchmark to determine if the model is producing
$\sim$100~kyr cycles.

\begin{figure}[ht]
  \centering
  \includegraphics[width=12cm]{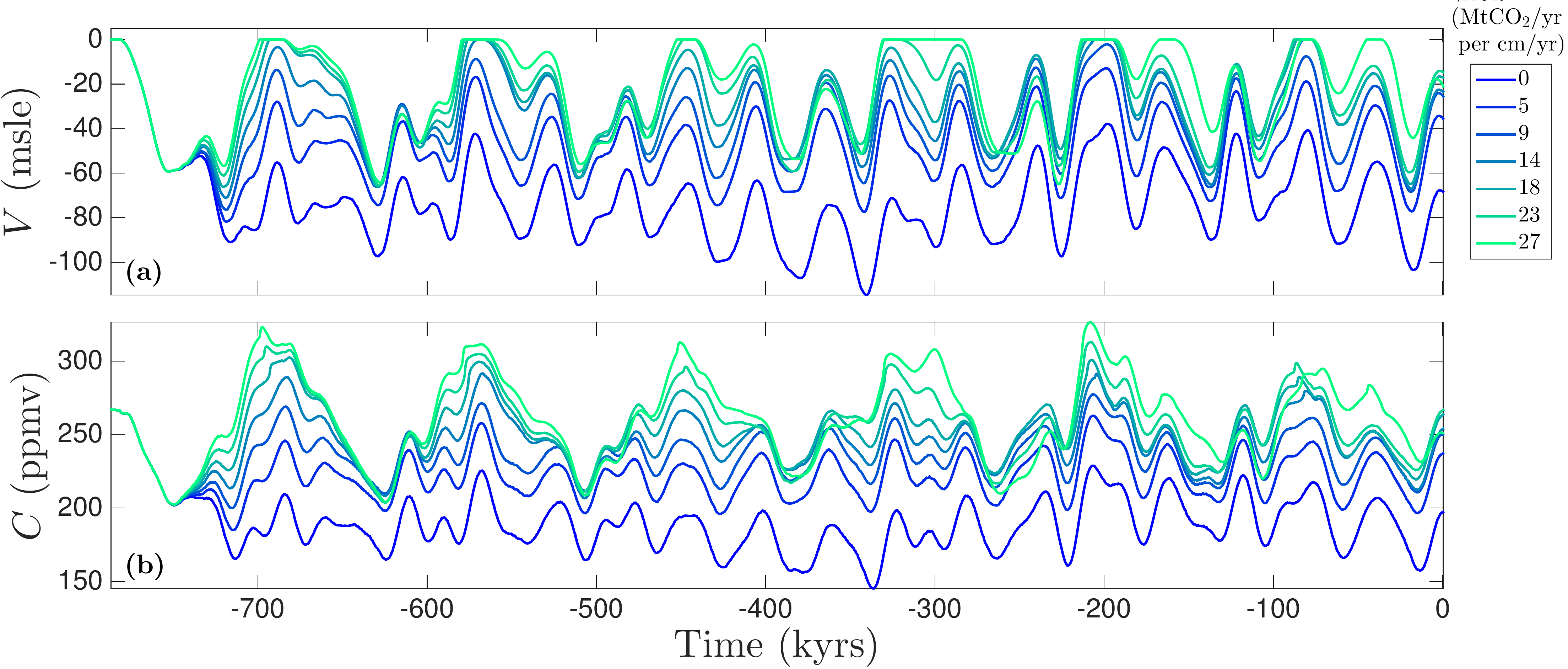}
  \caption{Increasing the sensitivity of MOR \cotwo{} emissions to sea
    level $\gMOR$, with $\gT=10$~ppmv/K, $\gSAV=18$ Mt\cotwo{}/yr per
    cm/yr change in sea level. Panel \textbf{(a)} is ice volume in metres sea level
    equivalent. Panel \textbf{(b)} is \cotwo{} concentration in the atmosphere.
    The change dominant period occurs in both $V$ and $C$ as $\gMOR$
    is increased, and occurs for the full time of the model run.}
  \label{fig:g2_range_620ka}
\end{figure}

Figure~\ref{fig:g2_range_620ka} shows how the full model varies with
increasing \gMOR{}.  For $\gMOR = 14$~Mt\cotwo{}/yr per cm/yr the
model has reasonable amplitude $C$ cycles ($\sim$80~ppmv) and
generates $V$ cycles with significant 100~kyr periodicity.  Thus the
amplitude of $C$ cycles are reasonably close to late-Pleistocene
values when the `full model' is close to replicating the ice core
$\sim$100~kyr glacial cycles (this trend holds across sensitivity
factor values).  Increasing $\gMOR$ increases the magnitude of $C$
cycles and adds greater 100~kyr variability.  We now consider the
periodicity of these model runs across a parameter sweep in the
model's three sensitivity factors \gT{}, \gSAV{}, and \gMOR{}.

\begin{figure}[ht]
  \centering
\includegraphics[width=12cm]{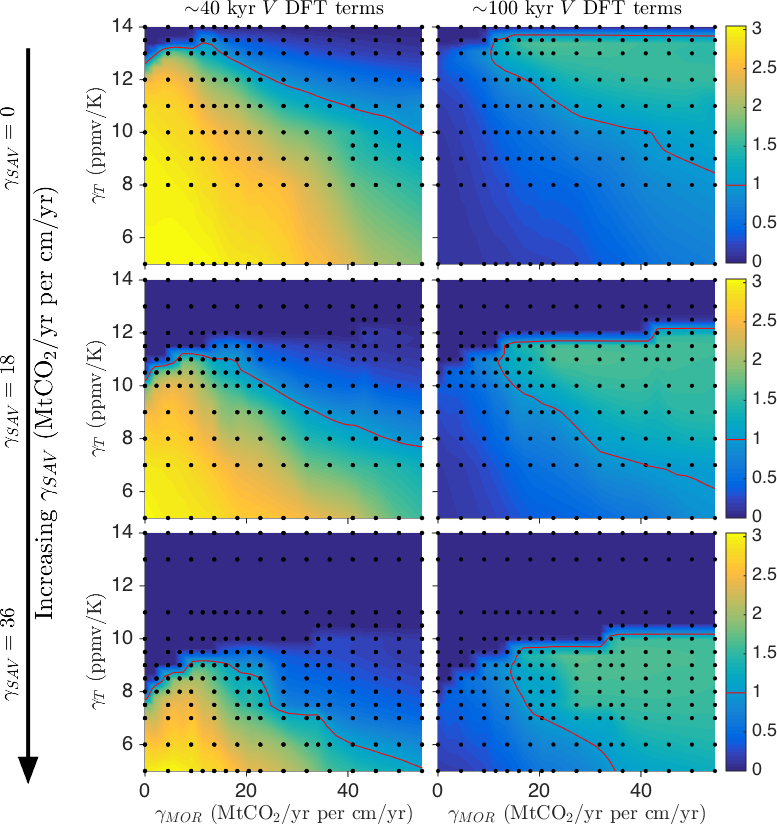}
\caption{ Periodicity of model runs for the last 787~kyrs with
  $\gSAV=0, 18, 36$ ~Mt\cotwo{}/yr per cm/yr change in sea level for
  the three rows. Plots show the strength of terms in the discrete
  Fourier transform of ice volume relative to the model output in
  figure~\ref{fig:IceCoreRep}(a). The left plots show 40~kyr terms,
  the right plots 80--120~kyr terms.  The red contour marks where the
  full model matches our modelled glacial replication.  Dark blue
  marks where the model enters a runaway glacial event, we chose not
  to plot these as FFTs are not valid for timeseries that have a sharp
  change in periodicity partway through. }
  \label{fig:ParamSweep}
\end{figure}

As mentioned above, we quantify the magnitude of the $40$~kyr and $\sim$100~kyr
periodicities in ice volume by comparing them to the same
periodicities (over the same time interval) in the model ice core
replication shown in figure~\ref{fig:IceCoreRep}c. Specifically, we
apply a discrete Fourier transform to each of these $V$ timeseries and
sum the terms in the frequency bands corresponding to 40~kyr and
80--120~kyr periodicity, then divide the `full model' value by the ice
core replication value --- if the result is above $1$ then there is
more power present (in that frequency band) in the full model than
there was in the calculated ice volume
for Late Pleistocene conditions.
This parity criterion is marked with a red contour line in
figure~\ref{fig:ParamSweep}. For $\sim$100~kyr cycles the minimum MOR
emissions sensitivity to meet this parity criterion is $\gMOR =
11$~Mt\cotwo{}/yr per cm/yr.

Figure~\ref{fig:ParamSweep} shows 40~kyr periodicity decreasing in
strength for increasing $\gMOR{}$, whilst the $\sim$100~kyr
periods increase in strength.  This matches the predictions in prior
sections and the behaviour in figure~\ref{fig:g2_range_620ka};  MOR \cotwo{}
emissions with a lag of $40$~kyrs oppose every second obliquity cycle
and create a stable feedback with an 80--120~kyr glacial cycle.

The trends in \gMOR{} values that cause the full model to reach and
exceed the parity criterion for $\sim$100~kyr $V$ cycles is as predicted
in prior sections.  Recall that MOR \cotwo{} emissions variations are
directly proportional to the magnitude of sea-level change, and
positive short-timescale intra-cycle feedbacks like \gT{} and \gSAV{}
increase sea-level change.  Therefore, the required \gMOR{} value to
match the parity criterion decreases as \gT{} or \gSAV{} increase.
When trading off between \gT{} and \gSAV{}, a lower \gSAV{} gives a
lower minimum \gMOR{} to reach parity for $\sim$100~kyr cycles, shown
by the top-right panel in figure~\ref{fig:ParamSweep} having the red
parity contour reach lower \gMOR{} values than in the lower-right
panel.

For very high $\gT$ or $\gSAV$, the $V$ cycle amplitude
increases. Runaway positive feedbacks in this limit (from larger ice
sheets and decreasing temperatures) lead to a permanent glaciation,
akin to a `Snowball Earth'.  It is not clear if such runaway scenarios
are reasonable representations of a marginal stability in the Earth
system, or a model failure (\ie{} parameterized feedbacks and forcings
becoming inaccurate in very cold, low \cotwo{} conditions that have no
parallel in the Pleistocene record).  The largest stable $\gT$ values
give model runs with sea-level changes of 85--100 metres, so our full
model captures glacial cycles with physical conditions similar to
historical glacial cycles.  Therefore we do not believe we are missing
parameter space relevant to the Pleistocene.

The power spectra for the full model at parity (\ie{} near the red
contour in figure~\ref{fig:ParamSweep}) have power in the
23/41/100~kyr bands of 3\%, 35\%, and 50\% respectively.  Compared to
our figure~\ref{fig:IceCoreRep} ice core replication (5.6\%, 23\%,
50\%), or sea-level data (4\%, 10\%, 55\%) the full model
is underpowered in the precessional band, and overpowered in the
obliquity band.  Despite this, the full model spectra (at parity) are
a reasonable match for glacial cycles.

Overall, the full model system can switch from 40~kyr glacial cycles
to $\sim$100~kyr cycles, the calculated $\sim$100~kyr cycles are
stable (figure~\ref{fig:g2_range_620ka}), and the minimum required
sensitivity of MOR \cotwo{} emissions to sea level is
$\gMOR=11$~Mt\cotwo{}/yr per cm/yr change in sea level
(figure~\ref{fig:ParamSweep}).  This $\gMOR$ requires MOR \cotwo{}
emissions at the upper end of a 95\% confidence interval (see
section~\ref{sec:discussion}) according to most prior work, thus this
$\gMOR$ value is possible, but not likely.

\clearpage{}

\section{Discussion}\label{sec:discussion}
We have presented a simplified model of climate through
glacial--interglacial cycles. The model comprises three variables ---
temperature, ice sheet volume, and \cotwo{} concentration in the
atmosphere --- these evolve according to equations based on the
physics of insolation, heat transfer in the atmosphere and Earth's
surface, radiative \cotwo{} forcing, ice flow under stress, proposed
MOR \cotwo{} emissions processes, and parameterizations of the
surface carbon system, subaerial volcanic \cotwo{} emissions, and
water vapour plus cloud forcing.  The
model calculates glacial-interglacial behaviour with insolation as the
sole driver of the system and \cotwo{} concentration in the atmosphere
as an internal feedback.  Although the model captures important and
fundamental physics, it neglects many processes that may affect the
results, which we discuss below.

    We treat the atmosphere as a single layer and parameterise the net
    upward and downward longwave greenhouse effects.  The
    parameterisation gives the overall energy balance between space,
    atmosphere, and ground but ignores changes in the internal
    atmospheric temperature structure. It is possible that important
    features are missed in this simplification, but our model does
    calculate reasonable present-day temperature distributions,
    seasonality (figure~\ref{fig:Seasons}), \cotwo{}-doubling scenarios
    (figure~\ref{fig:CO2_Doubling}), and glacial replications
    (figure~\ref{fig:IceCoreRep}).

    The ice model assumes a flat topography, distorted only by
    isostasy, and assumes no longitudinal variations in ice.  Flat,
    low-lying topography suppresses initial ice formation and ignores
    the complexity of advancing ice sheets across the terrain of North
    America and Europe, but figure~\ref{fig:IceCoreRep} shows our
    model replicating reconstructed sea-level timeseries, suggesting that
    the simplification is reasonable nonetheless. The computational
    complexity of the global 3D temperature and ice models required to
    relax these simplifying assumptions are too computationally
    expensive for Ma-scale studies; previous work on glacial ice
    sheets made similar simplifications \citep{Tarasov:1997vl,
      Fowler:2013dy}.

    We do not explicitly include oceans in our model, they are
    implicitly incorporated into the temperature-dependent surface
    system term in equation~\eqref{eq:Carbon_full} . However, oceanic
    effects (that we have neglected) should reduce volcanism-driven
    $C$ variations --- extra absorbtion/venting of \cotwo{} to/from
    oceans when the atmospheric \cotwo{} concentration is out of
    equilibrium with the surface ocean.  These are significant
    shortcomings, however there are no published ocean models that
    allow us to explicitly model oceanic \cotwo{} effects by
    dynamically replicating glacial-to-interglacial oceanic
    transitions.
    We believe the clear simplifications we make are better than
    building an ad-hoc ocean model.  It would be an improvement to the
    current work if the qualitative ideas of glacial oceanic changes
    (iron fertilisation of the South Atlantic, shifting latitudes of
    southern ocean westerlies, changing relative deepwater formation
    rates in the North Atlantic vs. Antarctica, etc...) were included in an
    ocean model that makes quantitative changes to atmospheric
    \cotwo{} concentration.

  We consider volcanic \cotwo{} emissions in our modelling, yet
  other glacially-driven volcanic changes could affect climate, such
  as SAV aerosols, MOR hydrothermal flux, and Fe flux.
  It is not clear if including these extra volcanic effects in a model
  would affect the switch to $\sim$100~kyr glacial cycles. If future
  research reveals any to have large climate feedbacks on 10's-of-kyr
  timescales, that would impact the conclusions of this work.

  Our model is a deterministic system and, unlike geological records
  of glacial cycles, has no noise on (\eg{}) $<$200~year timescales.
  However, noise does not effect our model's conclusions. When we
  introduce noise in input parameters, we see no change in qualitative model
  behaviour (appendix~\ref{sec:model-response-noise}).

  Even after accounting for simplifications, our model gives insight
  into glacial--interglacial behaviour.  Previous work takes dependent
  variables in the earth system (temperature, atmospheric \cotwo{}
  concentration, ice extent) and uses them as independent driving
  variables --- clearly shortcomings when considering the highly
  coupled glacial system whose key features emerge on the 10s-of-kyrs
  timescale.  This model addresses those features, with space for
  uncertainties to be reduced or further mechanisms explored.

We see a sharp distinction between climate feedbacks acting at
significantly less than the glacial period (short-timescale feedbacks)
and those acting at or above the glacial timescale.  Short-timescale
$C$ feedbacks are intracycle effects that modulate the magnitude of
each glacial, but because they carry little information from one
glacial cycle to the next, are ineffective at changing overall glacial periodicity.

Our model finds transitions from $40$~kyr cycles to $\sim$100~kyr
cycles as we increase MOR \cotwo{} emissions response to
rate-of-sea-level-change (\ie{} increasing $\gMOR$).  There is no
significant 100~kyr variability without the intercycle feedback from
MORs.  The transition mechanism is atmospheric \cotwo{} concentration
(influenced by MOR \cotwo{} emissions) acting to suppress a
glacial--interglacial transition triggered by insolation.  The
subsequent increase in sea-level periodicity from $40$~kyrs to
$80$--$120$~kyrs approximately doubles the magnitude of MOR \cotwo{}
variability (fig~\ref{fig:MORC_periodicSL}), and short-timescale $C$
feedbacks reinforce the new cycle and produce large $C$ changes that
dominate insolation such that only every second or third obliquity
cycle causes major deglaciation.

This transition mechanism inherently generates sawtooth patterns in
$V$ (fig~\ref{fig:varyingLag}b), describing a growing ice sheet,
interrupted growth (when $C$ and insolation are in opposition),
followed by further growth, and then a large deglaciation.

Our model's transition to $80$--$120$~kyr glacial cycles is broadly
consistent with the coupled oscillator model of
\cite{Huybers:2017vz}, suggesting analagous behaviour may govern our
system.

Under optimal conditions the model transitions to $\sim$100~kyr cycles
at $\gMOR=11$~Mt\cotwo{}/yr per cm/yr change in sea level.  Physically
this corresponds to MOR emissions of 91~Mt\cotwo{}/yr changing up to
$\pm 12\%$ across a glacial cycle, or (recalling that our \gMOR{} is
linear in baseline emissions and percentage change) a scaled
equivalent \eg{} 137~Mt\cotwo{}/yr changing up to $\pm 8\%$.  Are
these volcanic numbers feasible?  For our specified MOR lag time,
models predict $\gMOR=8\%$ \citep{Burley:2015ic} with little room for
error (uncertainties in the model inputs would not change predicted
\gMOR{} by $\pm 1$~percentage point), thus we must ask if
137~Mt\cotwo{}/yr is a reasonable background global volcanic \cotwo{}
emissions rate.

It is worth considering global MOR \cotwo{} emissions in some
detail, given the diverse literature.  There are two approaches to
estimating global MOR \cotwo{} flux, all based around measuring an
element that has a constant ratio to \cotwo{} in volcanic eruptions,
then using that fact (plus other assumptions) to calculate \cotwo{} emissions:
 \textit{1}) use the concentration
of  $^3$He in ocean water to infer the rate of MOR \cotwo{}
emissions.  The element must have a known lifetime in the ocean (preferably with no
non-volcanic inputs).  
\textit{2}) use the concentration of an element in volcanic rocks to
infer the \cotwo{} concentration in the source mantle.  Then apply a
melting fraction to generate an erupting mantle composition from the
source mantle, and multiply by the volume of mantle erupted per year
to calculate the rate of MOR \cotwo{} emissions.
The first approach has a single method, $^3$He in the oceans, which
has settled to values of 0--134~Mt\cotwo{}/yr \citep{Resing:2004ks}
and $18$--$176$~Mt\cotwo{}/yr \citep{Marty:1998vo} (2
std.dev.). Updated $^3$He flux values from \cite{Bianchi:2010hr} would
change these values to 0--101, and 9--93~Mt\cotwo{}/yr respectively.
For the second approach, the most recent work combining ratios of Nb,
Rb and Ba for melt inclusions calculates $18$--$141$~Mt\cotwo{}/yr (2
std.dev.)  \citep{LeVoyer:2017cb}.  Work using the undegassed
Siqueiros melt inclusions calculates $29$--$53$~Mt\cotwo{}/yr (2
std.dev.)  \citep{Saal:2002ks} (the Siqueiros melt inclusions may be
highly depleted, implying their derived global emissions value is an
underestimate) and volcanic glasses give $88$--$158$~Mt\cotwo{}/yr (2
std.dev.)  \citep{Michael:2015if}.
There could be systematic error in some of these measurements,
particularly given the sensitivity of the latter approaches to the
assumed average mantle melt fraction used to generate MOR basalts
(\ie{} erupting mantle composition) from the MOR mantle source
\citep{Cartigny:2008dz,Dasgupta:2010fw,LeVoyer:2017cb}.  Furthermore,
none of these studies include uncertainty in the degree of melting in
their random error, so errors are likely underestimated. Using the
latest melting models, \cite{Keller:2017cd} calculate a range of
53--110~Mt\cotwo{}/yr for \cotwo{} concentration in the MOR mantle source
is 100--200~ppmw. Extrapolating linearly (a vast simplification) to a 2-$\sigma$ range in
\cotwo{} concentration of 27--247~ppmw \citep{LeVoyer:2017cb} gives
14--135~Mt\cotwo{}/yr. Our required emissions of 137~Mt\cotwo{}/yr is
at the high end of the 95\% confidence interval for some of these
studies, therefore it is possible, although not likely, that the
global MOR \cotwo{} emissions rate is large enough to disrupt glacial
cycles, assuming no oceanic moderation of volcanic emissions.

However, if we assume volcanic \cotwo{} variability's effect on $C$ is
damped by oceanic absorption/emission, then the required MOR
parameters are outside the expected range.  This `oceanic damping'
logic is based on the idea that the surface ocean and atmospheric
\cotwo{} are in equilibrium, and that any attempt to change the
atmospheric \cotwo{} concentration is opposed by changes in ocean
chemistry. Such logic represents anthropogenic carbon changes well, but
glacial cycles probably involve changes in the physical ventilation of
the oceans, making the comparison inexact; modern models are a worst
case scenario.  Regardless, modern Earth system models
\citep{Archer:2009bm} suggest a factor of 4 increase in required
background MOR emissions --- necessitating 548~Mt\cotwo{}/yr, outside
the upper limits of MOR \cotwo{} emissions.  Even a factor of 1.5
increase would require unreasonable MOR \cotwo{} emissions.  Therefore
despite the uncertainty in oceanic \cotwo{}-damping effects,
background MOR \cotwo{} emissions are very unlikely to meet the
requirements for $\sim$100~kyr cycles after accounting for ocean
absorption.

The magnitude of changes in MOR and SAV \cotwo{} emissions are
proportional to the magnitude of sea-level change, and MOR \cotwo{}
emissions increase for longer period sea level changes.  Therefore, if
MOR \cotwo{} emissions are part of the transition mechanism from $40$
to $100$~kyr glacial cycles, the model suggests the following: 1)
transitioning to $\sim$100~kyr glacial cycles will increase the
magnitude of \cotwo{}, sea-level, and temperature changes ---
including warmer interglacial periods, and 2) a relatively large sea
level change should precede the transition to longer glacial cycles.

This process of volcanic \cotwo{} emissions altering glacial cycles is
consistent with the results of \cite{TzedakisPC:2017kn}, where the
summer insolation required to trigger full deglaciation increases
across -1.5~Ma to -0.6~Ma (after accounting for discount rate, whereby
deglaciation has a lower insolation threshold the longer an ice sheet
has existed).  Their discount rate is conceptually consistent with ice
sheet instability as explained in \cite{Clark:1998fx,AbeOuchi:2013hm},
however the changing insolation threshold is not readily explained by
existing theories.  A plausible explanation is a feedback cycle
whereby an increase in the magnitude of sea-level changes leads to
increased volcanic \cotwo{} emissions variability --- amplifying
\cotwo{} (and thus temperature) variations --- consequently amplifying
the next sea-level cycle. This eventually changes the period of
sea-level cycles as MOR \cotwo{} emissions variability becomes larger;
leading to further increases in MOR \cotwo{} emissions
(section~\ref{sec:mor-co_2-response}) and even larger sea-level
cycles, until the system reaches a new steady state with large, long
period sea-level cycles.  The feedback between volcanism and sea-level
would take several glacial cycles to reach a new equilibrium,
consistent with the 900~kyr transition time proposed in
\cite{TzedakisPC:2017kn}.

\section{Conclusion} \label{sec:conclusion}
We have presented a reduced-complexity model system that calculates the Earth climate
over the past 800~kyrs; a system with ice sheets, \cotwo{}
concentration in the atmosphere and other forcings evolving in
response to imposed insolation changes.  We demonstrated a match to
current planetary temperatures and GCM \cotwo{} doubling forecasts.
When driven with observed \cotwo{}, the model reproduces the 
 glacial sea-level record.  

 Our main research interest was quantifying the mid-ocean ridge (MOR)
 \cotwo{} emissions sensitivity to sea-level change necessary to
 induce $\sim$100~kyr glacial cycles, thus assessing the plausibility
 of volcanic mechanisms for creating an Earth system climate response
 not linearly related to insolation forcing.

Our model has no intrinsic 100~kyr variability until the lagged
response of MOR's \cotwo{} emissions to sea level change is included;
default behaviour is 40~kyr glacial cycles.
We calculate that MOR \cotwo{} variability, above a threshold
sensitivity to sea-level change, causes
glacial cycles at a multiple of insolation's 40~kyr obliquity cycle.
These $\sim$100~kyr cycles are asymmetric, and occur at both 80~kyr
and 120~kyr periods, replicating features of the late-Pleistocene
glacial record.

However, even under optimal conditions, $\sim$100~kyr cycles require
MORs' \cotwo{} emissions response be $11$~Mt\cotwo{}/yr per cm/yr rate
of sea-level change,
50$\%$ higher than the expected $7.3$~Mt\cotwo{}/yr
per cm/yr.  This requires background MOR \cotwo{} emissions of
137~Mt\cotwo{}/yr, within the 95\% confidence interval of (some)
estimates of MOR \cotwo{} flux.  However, under less optimal
conditions where oceanic effects moderate MOR emissions' effect on
$C$, required baseline MOR \cotwo{} emissions are over 200~Mt\cotwo{}
per year --- highly improbable.  This suggests that MOR \cotwo{}
emissions are not, in isolation, responsible for glacial cycles
$>40$~kyrs.

Of course, MOR \cotwo{} emissions do not act in isolation, and there
are relevant glacial mechanisms that do not operate in our model,
including regolith erosion\citep{Clark:1998fx}, secular \cotwo{}
decline \citep{Pagani:2010ks,Honisch:2009at}, and switching modes in
ocean ventilation
\citep{Franois:1997jy,Toggweiler:1999wj}.  These
mechanisms may interact with our existing processes to allow glacial
cycles at lower MOR \cotwo{} variability.  However, adding such
mechanisms would increase model complexity; furthermore, these
mechanisms are not precisely defined and would necessitate a wide
range of representative models and parameter sweeps.  Thus it is
unlikely that a mixed mechanism hypothesis for $\sim$100~kyr glacial
cycles can be tested until each mechanism is more precisely defined.

Our model system highlights other important features.
First, we calculate that the net changes in atmospheric \cotwo{}
concentration caused by MOR volcanism will approximately double when
sea-level periods increase from 40~kyrs to 100~kyrs. Therefore, if a
$\sim$100~kyr glacial cycle occurs, MOR volcanism acts to reinforce
that periodicity. 

Second, our model makes a distinction between intracycle and intercycle
feedbacks.  An intracycle feedback is a process with a timescale less
than half the glacial cycle period; therefore acting within a glacial
cycle.  Intracycle feedbacks affect the magnitude of glacial
cycles, but cannot change the glacial periodicity.  This result will
hold for any feedback process with constant sensitivity.

Third, we found that MOR systems with a 40~kyr lag between sea-level
change and consequent \cotwo{} emissions generate $100$~kyr cycles at
the lowest \gMOR{}. However, any intercycle feedback in the Earth
system can potentially generate $\sim$100~kyr cycles, and we calculate
significant power at $\sim$100~kyr for MOR lags of 30--80~kyrs.
Therefore the proposed volcanic mechanism for 100~kyr glacial cycles
is not dependent on a peculiarly specific MOR lag value (equivalently,
a particular mantle permeability).

Finally, without strong MOR \cotwo{} emissions sensitivity, our model
defaults to an obliquity-linked glacial cycle with a $40$~kyr period;
precession's $23$~kyr cycle has little effect on the ice sheet.  This
result is in agreement with previous work considering integrated
summer forcing, and is the first time that $40$~kyr response has been
shown dominant in a model with radiative feedbacks.  Therefore our
model opposes the hypothesis that precession-linked glacial cycles may
have occurred before the mid-Pleistocene transition, with anti-phase
changes in Antarctic and Greenland ice mass at the $23$~kyr period
leaving a predominant 41~kyr signal in the $\delta^{18}\textrm{O}$
record \citep{Raymo:2006ca}.

The model's conclusion could be sensitive to some of our
simplifications, such as the oceans' interaction with volcanic
emissions on glacial timescales and the climate effect of other
variable volcanic elements.  However, these effects are all beyond
current understanding and it is hard to predict their effect on our
model.  Complete understanding of glacial cycle dynamics will require
models including several of the mechanisms currently proposed in
literature. We hope that future work can build on the base that we
have presented.

\paragraph{Acknowledgements} The research leading to these results has
received funding from the European Research Council under the European
Union's Seventh Framework Programme (FP7/2007-2013) / ERC grant
agreement number 279925. The University of Oxford Advanced Research
Computing (ARC) facility was used in this work
(doi:10.5281/zenodo.22558).  Katz thanks the Leverhulme Trust for
additional support. We thank D. Battisti and J. Moore for helpful discussions.

\appendix{}
\section{Appendix}\label{sec:appendix}
\subsection{EBM Model} \label{sec:ebm-model}
The EBM is constructed around energy balance equations for the middle
atmosphere \eqref{eq:EBM_atm}, ground surface \eqref{eq:EBM_surf}, and
subsurface \eqref{eq:EBM_subsurf}, repeated below:
\begin{align}
c_a \pdiff{T_a}{t} = S_a + I_a + F_s + D_a \;\;,\label{eq:EBM_atm_apdx}\\
c_s \pdiff{T_s}{t} = S_s + I_s - F_s + F_{ss} \;\;,\label{eq:EBM_surf_apdx}\\
c_{ss} \pdiff{T_{ss}}{t} = - F_{ss} \;\;,\label{eq:EBM_subsurf_apdx}
\end{align} 
where $a, s, ss$ subscripts define atmospheric, surface and subsurface
quantities. $c$ is heat capacity (Jm$^{-2}$K$^{-1}$), $S$ is the solar
radiation (shortwave), $I$ is infrared longwave radiation, $F$ is the
sensible heat flux, and $D_a$ is the meridional heat flux.  All the RHS
quantities are Wm$^{-2}$ and are detailed below.

Solar radiation is treated as reflecting between the ground and a
single atmospheric layer.  The atmosphere has reflectivity $R$,
absorption $A$, and transmissivity $T$ --- these sum to 1.
The ground has reflectivity (equivalently, albedo) $\alpha$, which
has two values representing ice/non-ice conditions. Thus
\begin{align}
S_a = AS +\frac{TAS\alpha}{1-\alpha R} \;\;,\\
S_s = TS \frac{1-\alpha}{1-\alpha R} \;\;,
\end{align} 
where the ground-atmosphere reflections are included as the
sum of a geometric series.
The model atmosphere has a single values for shortwave reflectivity
and transmissivity, whereas real atmospheric reflectivity should vary
with latitude due to increased cloud cover at high latitudes
\citep{Donohoe:2011cr}.  Higher reflectivity at high latitude would 1)
make variable insolation a weaker driver of glacial cycles, and 2)
reduce the albedo effect of ice sheets, which are stronger in our
model than in GCMs.

The infrared components are treated as imperfect black body radiators,
giving
\begin{align}
I_s = \epsilon_a \sigma T^4_{as} -\sigma T^4_{s} \;\;,\\
I_a = \sigma T^4_{s} - \epsilon_a \sigma  T^4_{as} - \epsilon_a \sigma
T^4_{ul} \;\;,
\end{align}
where $\epsilon_a $ is the longwave atmospheric emissivity, $\sigma$
is the Stefan-Boltzmann constant, $T_{as}$
is the atmospheric temperature at the ground surface and $T_{ul}$ at
upper atmosphere.  The atmospheric temperatures are related to the
middle atmosphere temperature by a constant moist adiabatic lapse rate
$\Gamma_m$ of $6.5~$K/km.  The alternative of a spatially and
temporally varying lapse rate requires assumptions about the
global hydrological cycle that we choose to circumvent.  Applying the lapse
rate to the previous equations gives
\begin{align}
I_s = \epsilon_a \sigma T^4_{as} -\sigma (T_a+\Gamma_m H_{as})^4 \;\;,\\
I_a = \sigma T^4_{s} - \epsilon_a \sigma  (T_a+\Gamma_m H_{as})^4 - \epsilon_a \sigma
(T_a+\Gamma_m H_{ul})^4 \;\;,\label{eq:I_a}
\end{align}
where $H_{ul}$ is the distance from the middle atmosphere to the upper
atmosphere, $H_{as}$ is the distance from the middle atmosphere to the
ground surface, which varies with thickening ice sheets and bed
depression. The upper atmosphere is defined as the height the
atmosphere becomes optically thin to IR radiation and varies with $C$
to represent radiative \cotwo{} forcing (see
section~\ref{sec:adding-co2-forcing}).

Sensible heat flux scales linearly with the temperature difference
between adjacent surfaces:
\begin{align}
F_{ss}= K_s \left( T_{ss} - T_s   \right) \;\;,\\
F_s = K_{ss} \left( T_s - (T_a+\Gamma_m H_{as})  \right) \;\;,
\end{align} 
where $K$ is the coefficient parameterising the heat flux.

The meridional heat transport scales with the meridional heat gradient,
\begin{align}
D_a = \pdiff{}{\phi}\left( -  K_a \left|\pdiff{T_a}{\phi} \right| \pdiff{T_a}{\phi} \right) \;\;,
\end{align}
and this transport is tuned to a reasonable value by selecting $K_a$.
We follow the $1000$~J/K per degree latitude used in \cite{Huybers:2008fg}.

To get to the energy available for melting ice the energy balance
model is run for a year, with 1-day explicit timesteps to account for
seasonality.  At each timestep, the energy flux in/out of the ground
surface is calculated; this energy flux changes the surface
temperature or melts ice according to the local conditions. The
thickness of ice melted is then calculated assuming constant ice
density on the ice sheet's upper surface.

Finally, the model needs precipitation to accumulate ice.
Precipitation is dependent on temperature as described in
eqn~\eqref{eq:Precip}.  At each timestep, wherever the atmospheric
temperature at the ground surface is below freezing, this
precipitation falls as snow and creates an ice sheet.  Where atmosphere
is above freezing, the precipitation has no effect on the ice
thickness or energy balance.

\begin{table}[h!]
  \begin{center}
    \begin{tabular}{lrll}
      \hline    
      Variable		&	Value	&	&	Description					\\
      \hline
$\;$$\alpha_g$	&	$0.3$		&	&land albedo \\
$\;$$\alpha_i$	&	$0.8$		&	&ice albedo\\
$\;$$\Gamma_m$	&	$6.5$		&K/km	&moist adiabatic lapse rate \\
$\;$$\epsilon_a$	&	$0.85$		&	&longwave atmospheric emissivity \\
$\;$$\rho_l$		&	$900$	&kg/m$^3$	&ice density\\
$\;$$\rho_w{}$	&	$1000$	&kg/m$^3$	&water density\\
$\;$$\rho_a$		&	$1.5$	&kg/m$^3$	&surface air density\\
$\;$$\sigma{}$	&	$5.67\times 10^{-8}$	&W/(m$^2$K$^4$)	&Stefan-Boltzmann constant \\
$\;$$C_p$		&	$2100$			&J/(kg K)	&specific heat capacity of water \\
$\;$$C_{air}$		&	$1.5$			&J/(kg K)	&specific heat capacity of air \\
$\;$$C_{a}$		&	$5000 \rho_a C_{air}$&J/(m$^2$K)	&atmospheric heat capacity \\
$\;$$C_{s}$		&	$5\rho_iC_p$		&J/(m$^2$K)&surface heat capacity\\
$\;$$C_{ss}$		&	$10\rho_iC_p$		&J/(m$^2$K)&subsurface heat capacity\\
$\;$$g$			&	$9.8$	&m/s$^2$       &gravitational acceleration \\
$\;$$H_{as}$		&	$5$			&km	&height of middle atmosphere above sea-level\\
$\;$$H_{ul}$		&	$2$			&km	&upper atmosphere thickness\\
$\;$$K_s$		&	$5$		&W/(m$^2$ K) &Thermal transmittance, surface--atmosphere,\\
 & & & the sensible heat flux coefficient \\
$\;$$K_{ss}$		&	$2$		&W/(m$^2$ K)	&Thermal transmittance, surface--subsurface \\
$\;$$K_a$		&	$1000/(^o$ lat)&J/K	& meridional heat flux coefficient\\
$\;$$L_v$		&	$2.5 \times 10^6$		&J/kg	&latent heat of vaporization\\
$\;$$L_m$		&	$3.34 \times 10^6$		&J/kg	&latent heat of melting\\
$\;$$L_s$		&	$2.84 \times 10^6$		&J/kg	&latent heat of sublimation \\
$\;$$S$			&	$1365$		&W/m$^2$	&solar constant \\
$\;$$A$			&	$0.2$		&	&absorption of atmosphere \\
$\;$$R$			&	$0.3$		&	&reflection of atmosphere\\
$\;$$T$			&	$0.5$		&	&transmission of atmosphere\\
      \hline
    \end{tabular}
  \end{center}
  \caption{Parameters used for the EBM.}
  \label{Tab:EBM_Parameters}
\end{table}

\subsection{EBM Modifications} \label{sec:ebm-modifications}
We make two major changes to the EBM scheme of \cite{Huybers:2008fg}:
adding longwave radiative forcings based on atmospheric composition,
and creating a variable precipitation based on local temperature.
These are explained below.

First we discuss our longwave radiative modifications;   
adding forcings representing \cotwo{}, water vapour, lapse rate,
and cloud effects.  These radiative forcings are treated with two
terms: one for the \cotwo{} forcing, and another for the aggregate
effects of water vapour, lapse rate and cloud forcings.  We add these
forcings to a single-layer atmosphere that has an imposed linear
temperature profile $\textrm{d}T/\textrm{d}z$ of $6.5$~K/km. The base
of the atmosphere is at the ground surface (land or the top of an ice
sheet).  The upper surface of the atmosphere is the mean height at
which the atmosphere becomes transparent to longwave radiation and
allows longwave emissions to to escape to space.  In the model's
neutral state, this is 6.5~km above sea level, well within the
troposphere's linear temperature profile. Therefore, changes in
the height of the mean emissions layer will change its temperature;
this changes the power of longwave emissions to space according to the
Stefan-Boltzmann law.

Greenhouse gas forcing occurs because such gases alter the height at
which atmosphere becomes transparent to longwave radiation. Our model
scheme is guided by this physics; implementing top-of-atmosphere
forcing by altering the height of the emissions layer.

The upward longwave (ULW) radiative \cotwo{} forcing is derived from
 \cite{Myhre:1998tg} global mean forcing,
\begin{align} \bar{R}_{\textrm{ULW}} = \lambda \log_2\left(
    \frac{C}{C_{0}}\right) \;\;,
      \label{eq:ULW}
    \end{align} 
    where $\bar{R}_{\textrm{ULW}}$ is the global average radiative
    forcing (we use $R$ instead of typical forcing
    terminology `$F$' to avoid confusion with the EBM's heat flux
    terms), $\lambda$ is \cotwo{} sensitivity in W/m$^2$ per
    \cotwo{}-doubling, and $C_0 = 280$~ppmv is the reference
    concentration of \cotwo{} in the atmosphere.  We follow the
    conclusions of \cite{Stap:2014bs} and \cite{Kohler:2010db} for
    greenhouse forcing in the Pleistocene and assume a $30$\% increase
    in \cotwo{} forcing magnitude from syncronous increases other
    greenhouse gases. Thus $\lambda$ is $4.85$~W/m$^2$ per
    \cotwo{}-doubling.

    We convert this global average forcing to a latitude-dependent
    forcing ${R}_{\textrm{ULW}}$ by changing the height (and therefore
    temperature) of the atmospheric layer emitting longwave radiation
    to space --- details are given in
    appendix~\ref{sec:adding-co2-forcing}.  A reasonable summary is to
    consider a global temperature change of the emitting layer $\Delta
    T$ (assumes $\textrm{d}T/\textrm{d}z$ independent of $\phi$),
    changing Stefan-Boltzmann emissions by $(T_{ul} + \Delta T)^4 - T_{ul}^4$.
    Upper atmosphere temperature $T_{ul}$ varies with $\phi$, and the forcing
    is thus latitude-dependent.  This forcing is applied to the net
    infrared energy balance of the atmosphere in equation~\ref{eq:I_a_full} below.

    DLW is parameterized to match the
    calculations in \cite{Cai:2012cm} of longwave \cotwo{} forcing at
    the bottom of the atmosphere in a 2D GCM.  Our parameterised
    equation is a logistic function and an exponential
    \begin{align}
      R_{\textrm{DLW}} = K \left( \frac{1}{1+e^{(a-\phi)/b}}  +
        \frac{e^{(-\phi/c)}}{3}  \right) \;\;,
 \label{eq:DLW}
   \end{align}
   where $K$ is the maximum forcing, $a=30^{\textrm{o}}$ is the latitude
   of the logistic function turning point, $b$ scales the width of the
   logistic function's growth region and $c$ scales the e-fold length
   of the decreasing exponential. For a given \cotwo{} concentration,
   $K$ is equal to the maximum value of ${R}_{\textrm{ULW}}$.

   We model the combined global energy balance effects of water
   vapour, lapse rate and clouds as a radiative longwave forcing,
   varying linearly with changes in $\bar{T_a}$ from preindustrial
   conditions.  This combined forcing is a parameterisation such that
   our model matches the climate sensitivity of general circulation
   models for a \cotwo{} doubling from preindustrial conditions ---
   details are given in section~\ref{sec:adding-WLC-forcing}.  We
   extrapolate from \cotwo{} doubling experiments and assume that the
   net water vapour, lapse rate, and cloud feedback is also linear for
   glacial--interglacial climate changes, literature does not yet have
   significant statistical proof or opposition to this assertion
   \citep{Bony:2015fd,Braconnot:2015hd,Harrison:2015ej,Hopcroft:2015ur,Harrison:2016uo}.
   Having defined our radiative forcings, we now incorporate them into
   the EBM longwave balance.

The net longwave radiation balance of the atmosphere $I_a$ has three terms
representing, respectively, the longwave emissions from the ground
(absorbed by the atmosphere), emissions from the atmosphere to the
ground, and emissions from the atmosphere into space.  Applying the
collective radiative forcings to $I_a$ gives
\begin{align}
I_a = \sigma T^4_{s} &- \big(\epsilon_a \sigma  \big(T_a-\Gamma_m
H_{as}\big)^4 +  R_{\textrm{DLW}}\big)... \nonumber \\
 &-\:\: \epsilon_a \sigma
\big(T_a+\Gamma_m (H_{ul}+\Delta z_{C} + \Delta z_{\textrm{WLC}})\big)^4 \;\;,\label{eq:I_a_full}
\end{align}
where $\sigma$ is the Stefan-Boltzmann constant, $\epsilon_a$
is atmospheric emissivity, $T_a$ is the temperature of the middle
atmosphere, $\Gamma_m$ is the temperature profile in the atmosphere
$\textrm{d}T/\textrm{d}z$, $H_{as}$ is the middle-atmosphere-to-surface
height, $H_{ul}$ is the default middle-atmosphere-to-upper-layer
height, $\Delta z_{C}$ is the change in upper layer height due to
\cotwo{} concentration in the atmosphere, and $\Delta
z_{\textrm{WLC}}$ the change in upper layer height modelling the
parameterised water vapour, lapse rate, and cloud feebacks.

Finally, we consider the EBM's precipitation model. Precipitation rate
is used to track snowfall in the EBM, with snow falling when the
atmosphere at ground level is below $0^{\circ}$C (precipitation above
$0^{\circ}$C is ignored). 

We change the \cite{Huybers:2008fg} model's fixed 1~m/yr precipitation
rate to a temperature-dependent precipitation rate, calculated at each
timestep and gridcell in the EBM.  This more realistic precipitation
model was introduced to limit the growth of very large ice sheets
(equivalent to $>150$~m sea-level change) that occured under the fixed
precipitation rate, driven by unrealistically high 1~m/yr snowfall on
3~km thick ice sheets at 70-80N.  Our precipitation model is
\begin{align} 
 P = 
  \begin{cases} 
   P_c e^{\left(\frac{_{~}T_{~}}{20} - 1 \right)} & \textrm{if } \phi \geq 50\deg \\
   P_c e^{\left(\frac{T_{50}}{20} - 1 \right)} & \textrm{if } \phi < 50\deg
  \end{cases}
 \;\;,
      \label{eq:Precip}
\end{align}
where $P$ is precipitation in m/yr, $P_c = 2$ m/yr is a scaling
constant, $T$ is surface air temperature in Celsius and $T_{50}$ is
temperature at 50\deg{} latitude.  The $T/20$ scaling and $P_c$ value
are derived from an exponential best-fit to the ERA-interim reanalysis
product (D. Battisti -- pers. comm.). Physically, this exponential
parameterisation represents the decreasing vapour saturation of colder
air leading to reduced precipitation \citep{Pierrehumbert:2007wg},
although we skip the details of why air remains near vapour saturation
and the complexity of tracking moisture from source to sink.
Using $T_{50}$ to determine precipitation below 50\deg{} enforces
mid-latitude deserts. What about the tropics? Across
glacial-to-interglacial conditions our model has lowest-latitude
snowfall occurring between 38-48N;  thus the parameterisation does not
artifically reduce tropical snowfall --- the tropics are
snow-free even in glacial conditions.

Equation~\eqref{eq:Precip} has the consequence that
thick, cold ice sheets become drier during a glacial period, as expected.

\subsection{Calculating Radiative CO$_2$
  Forcing} \label{sec:adding-co2-forcing}
The reference model for the radiative \cotwo{} effect uses a
combination of line-by-line, narrow-band and broad-band radiative
transfer schemes \citep{Myhre:1998tg} that produces the simple
parameterisation of equation~\eqref{eq:ULW} with a logarithmic
scaling.  Here we describe how this
radiative forcing is incorporated into our energy balance model.

Equation~\eqref{eq:ULW}'s forcing $\bar{R}_{\textrm{ULW}} $ is a
planetary average, if applied uniformly it would overestimate
radiative \cotwo{} forcing at the poles and underestimate near the
equator.  We instead calculate the change in the average temperature
of the emitting layer of the atmosphere $\Delta \bar{T}$ required to
match $\bar{R}_{\textrm{ULW}} $. Then we convert $\Delta \bar{T}$ to
an equivalent change in height of the emitting layer (\ie{} the height
above which the atmosphere is optically thin to IR radiation).  This
is consistent with the physics of higher \cotwo{} concentrations
making the atmosphere optically thicker to IR radiation.

This is derived below, beginning with a perturbation to a 
default state `0'.
\begin{align}
F_{0} +\bar{R}_{\textrm{ULW}}  & =  \epsilon_{a} \sigma ( T_{0} + \Delta \bar{T})^{4} \;\;,\\
F_{0} +  \lambda \, \textrm{Ln}(C/C_{0})  &=  \epsilon_{a} \sigma (
T_{0} + \Delta \bar{T})^{4} \;\;,
\end{align}
where $F_0$ is the unperturbed longwave radiative flux to space, $T_0$
is the mean temperature of the emitting layer of the atmosphere, and
$\bar{R}_{\textrm{ULW}} $ is from
eqn~\eqref{eq:ULW}. Thus the perturbation is:
\begin{align}
\lambda \, \textrm{Ln}(C/C_{0})  &=  \epsilon_{a} \sigma
\left(     4T_{0}^{3} \Delta \bar{T}  +  6T_{0}^{2} \Delta \bar{T}^{2}  +  4T_{0} \Delta \bar{T}^{3}  +  \Delta \bar{T}^{4} \right) \;\;. \label{eq:DFtoAtmH_step1}
\end{align}
Equation~\eqref{eq:DFtoAtmH_step1} is fourth order in $\Delta{} \bar{T}$ and thus
computationally expensive.  However, $\Delta \bar{T} \ll 0.1\bar{T}$ thus terms
above second order change the RHS by less than 1\%, and we
instead solve a quadratic in $\Delta \bar{T}$. Thus
equation~\eqref{eq:DFtoAtmH_step1} becomes:
\begin{align}
\Delta \bar{T} &= \frac{-T_{0}^{2} \pm \sqrt{ T_{0}^{4} -  \frac{3
    \lambda}{2\epsilon_{a} \sigma}  \textrm{Ln}(C/C_{0})   } }{3
T_{0}}  \;\;.
\end{align}
Finally, using the linear temperature profile $\Delta \bar{T} = - \Gamma_{m}
\Delta z_{C}$, discarding the nonsensical solution and rearranging
\begin{align}
\Delta z_{C} &= \frac{T_{0}}{3\Gamma_{m}} \left(1 - \sqrt{1 -\frac{3\lambda}
    {2\epsilon_{a} \sigma T_{0}^4}  \textrm{Ln}(C/C_{0}) }\;\right)   \;\;, \label{eq:z_vs_T} 
\end{align}
where $\Delta z_{C}$ is the change in height of the emissive layer of the
atmosphere and is constant with respect to latitude.  To illustrate
how equation~\eqref{eq:z_vs_T} relates to climate: if $C>C_0$ then
$\textrm{Ln}(C/C_{0}) $ is positive, the square root is a small
positive number, and $\Delta z_{C}$ is positive --- thus the emitting
layer becomes higher and colder, reducing radiative emissions to
space, and forcing the planet to warm to re-establish equilibrium (as
expected for an increase in \cotwo{}).

\subsection{Calculating Other Radiative Forcing} \label{sec:adding-WLC-forcing}
Over glacial--interglacial cycles \cotwo{} forcing is not the only
significant radiative effect, there are contributions from water
vapour, lapse rate, and clouds (subsequently WLC).  Unlike \cotwo{}
forcing, the WLC contributions do not have well-established, simple
parameterised equations (see equation~\eqref{eq:ULW}) to predict their
effects.  In this section we discuss how WLC forcings are
incorporated into our model.

The most detailed modelling of WLC effects on planetary temperature
comes from general circulation models forecasting the global warming over the next
100 years, and we can extract a parameterisation from these models.
Parameterisation is required because the actual water vapour, lapse
rate, and cloud schemes in GCMs are too computationally
expensive for our model goals.  Fortunately these models' results can
be compiled into the following equation, assuming each mechanism
operates as an independent linear perturbation feedback on average planetary
temperature.

\begin{align}
\Delta \bar{T_a} &= \lambda_0 \Delta \bar{R}_{\textrm{ULW}} +
f_{\textrm{net}} \Delta \bar{T_a} \;\;, \\
\Delta \bar{T_a} &= \lambda_0 \Delta \bar{R}_{\textrm{ULW}} + f_{\textrm{wv}} \Delta \bar{T_a} +
f_{\textrm{lr}} \Delta \bar{T_a} + f_{\textrm{c}} \Delta \bar{T_a} +
f_{\textrm{a}} \Delta \bar{T_a}   \;\;, \label{eq:LinGCM_feedbacks}
\end{align}
where $\Delta$ denotes a change in a quantity, $f_x$ are unitless
feedback parameters operating: $ f_{\textrm{net}}$ in total,
$f_{\textrm{wv}}$ from water vapour, $f_{\textrm{lr}}$ from lapse rate
, $f_{\textrm{c}}$ from clouds, and $f_{\textrm{a}}$ from surface ice
albedo. ${R}_{\textrm{ULW}}$ and $T_a$ are as defined previously.

Using
equation~\ref{eq:LinGCM_feedbacks} GCMs show $f_{\textrm{net}}$
range 0.49--0.73.  Our model's $\lambda_0 \Delta
\bar{R}_{\textrm{ULW}}$ and $f_{\textrm{a}}$ under modern insolation
conditions comparable to GCM runs  are 1.08~K per \cotwo{} doubling and
0.273 respectively. With these we calculate the cumulative water
vapour, lapse rate, and cloud feedbacks $f_{\textrm{WLC}}$ required
for our model to be in line with GCM $f_{\textrm{net}}$ values using $
f_{\textrm{net}}= f_{\textrm{WLC}} + f_{\textrm{a}}$.  This gives a $f_{\textrm{WLC}}$
range 0.22--0.46.

Now we consider how to incorporate this WLC feedback in our
model. $f_{\textrm{WLC}}$ cannot be directly used in our EBM
because it describes the equilibrium state --- the increase in
equilibrium average planet temperature from WLC effects per Kelvin
increase in average planet temperature from other effects --- rather
than a forcing that causes the equilbrium state.  We calculate
a longwave radiative forcing that will have the equivalent effect on
equilibrium temperature $\bar{R}_{\textrm{WLC}} =
f_{\textrm{WLC}}/\lambda_0$; this global average radiative forcing is
included in the model by the same procedure described in
section~\ref{sec:adding-co2-forcing}.  This method assumes WLC
radiative effects can be linearly superimposed with longwave \cotwo{}
forcing.  Having chosen the method, we must select a
$f_{\textrm{WLC}}$ value from within the available range.

We select a value of $f_{\textrm{WLC}}=0.42$ to optimise the model's
replication of the sea level record when forced with ice core \cotwo{}
values (see figure~\ref{fig:IceCoreRep});  the model's overall
feedback is 0.69, at the upper end of the GCMs' range.

To summarise, our WLC parameterisation creates a mean global
forcing proportional to the current deviation from mean preindustrial
temperature.  The mean global forcing in this parameterisation is
chosen to match the climate sensitivity of GCMs for \cotwo{}-doubling
experiments, giving us high confidence of
accuracy on this metric.  

However, we extrapolate assuming the forcing is linear with respect to
temperature across the full glacial--interglacial climate variability;
literature does not yet have significant statistical proof or
opposition to this assertion
\citep{Bony:2015fd,Braconnot:2015hd,Harrison:2015ej,Hopcroft:2015ur,Harrison:2016uo}. The
good match we obtain to sea-level records in
figure~\ref{fig:IceCoreRep} supports this assertion.

The WLC forcing is applied by changing the mean height of longwave
emissions to space. Consequently, WLC forcing has the same latitudinal
pattern as \cotwo{} forcing. Is this accurate for each of the three
components?

For water vapour it is probably
    reasonably accurate; the mechanism of changing atmospheric
    concentration affecting the longwave optical thickness of the
    atmosphere applies to both \cotwo{} and H$_2$O.  Although H$_2$O
    has a less homogenous distribution across latitudes, which may
    introduce some differences.

Lapse rate
    affects longwave emissions by changing $\textrm{d}T /\textrm{d}z$,
    causing a forcing equivalent to changing the height of the
    emissions layer. Assuming that the change in lapse rate is
    latitude independent, the  lapse rate effect is physically
    consistent with WLC forcing .  However, it is not clear that
    changes in lapse rate will be latitude independent.  

    Cloud forcing is the least understood of these feedbacks
    \citep{Soden:2006gg,IPCC:2013_PhysSci}, but it is likely that high
    latitudes experience greater cloud forcing feedbacks
    \citep{Cronin:2015hl}, unlike our WLC forcing.  Furthermore,
    clouds affect both the longwave and shortwave radiation balance,
    and we only represent a longwave effect.  A cloud model
    calculating changes to the atmospheric shortwave parameters would
    interact with insolation (\eg{} an increase in low-altitude
    high-latitude clouds increases shortwave reflectivity, therefore
    changes in insolation would have a reduced effect at high
    latitude); potentially creating emergent behaviours that are not
    captured our model.  The redeeming feature of our cloud
    parameterisation is that it matches the average global cloud
    forcing of a GCM cohort.

We believe our WLC forcing model represents radiative water vapour
forcing and lapse rate forcing well, but cloud forcing poorly.  In the
\cite{Soden:2006gg} model comparison, water
vapour, lapse rate and cloud feedbacks are, respectively,  1.8, -0.84, and 0.68
Wm$^{-2}$K$^{-1}$.  Therefore we capture the effect of the largest
forcing contributions.  We believe this represents a reasonable
approximation for our low-complexity modelling framework.

\subsection{Ice Sheet Model} \label{sec:ice-sheet-model-appdx}
Our ice sheet model follows that of \cite{Huybers:2008fg}. However,
for completeness, we state the model equations and parameter values
here. We begin with a PDE for ice thickness deriving from conservation of mass
\begin{align}
  \pdiff{h}{t} &= B - \pdiff{}{x}\left( \bar{u}h \right)
  \;\;, \label{eq:IceContinuity_simple}
\end{align}
where $B$ is the net accumulation rate (precipitation minus melting)
in metres of ice per second, $\bar{u}$ is the depth-averaged ice
velocity, and $x$ is distance along a line of latitude.  Combining
eqn.~\eqref{eq:IceContinuity_simple} with the shallow-ice conservation
of momentum equation, Glen's Law, and a variable bed height gives
 \begin{align}
\pdiff{h}{t} &= B + \pdiff{}{x}\left( \frac{2A(-\rho g)^n}{n+2} \left|
    \pdiff{(h+H)}{x} \right| \pdiff{(h+H)}{x} (H+h)^{n+2} +
    u_bh\right)\;\;,
 \label{eq:Ice_dhdt_FULL}
   \end{align}
   where $A$ is the ice deformability constant in Pa$^{-3}$s$^{-1}$,
   $\rho$ is ice density, $g$ is gravitational acceleration, $n$ is
   the exponent relating stress to strain in Glen's law, $H$ is the
   height of the ground surface (see eqn.~\eqref{eq:BedRelaxation})
   and $u_{b}$ is the horizontal (sliding) speed of the base of the ice
   sheet. The sliding speed is calculated by

\begin{align}
 u_b &= \frac{2D_o a}{(m+1)b}
\left( \frac{\rho_i g h \left|\pdiff{h}{x}\right|}{2D_o\mu}
\right)^{\!\!\!m} \! \cdot
 \left[1-\left(1-\frac{b}{a}\textrm{min}\Big(h_s , \frac{a}{b}\Big)\right)^{m+1}\right]\;\;, \label{eq:BasalSliding_Full}
\end{align}
where $D_o$ is the reference sediment deformation rate, $m$ is the exponent in
the stress-strain relationship, $\rho_i$ is ice density, $\mu$ is the
sediment reference viscosity, $a= \rho_i g h | \partial h / \partial x
|$, the shear stress imparted to the sediment by ice flow above it,
and $b=g(\rho_s - \rho_w)\textrm{tan}(\phi_s)$ the rate of increase of
shear strength with depth in sediment.  Parameter values are given in
table~\ref{Tab:Ice_Parameters}.

Finally, the bed height variation is calculated as a local relaxation
to isostatic equilibrium
\begin{align}
\pdiff{H}{t} = \frac{1}{T_b} \left(  H_{eq} - H - \frac{\rho_i
    h}{\rho_b}  \right) \;\;, \label{eq:BedRelaxation}
\end{align}
where $T_b$ is the bed relaxation time constant, $H_{eq}$ is the
equilibrium bed height, $\rho_i$ is ice density, $\rho_b$ is the
bedrock density.  

   Eqn.~\eqref{eq:Ice_dhdt_FULL} is solved with a semi-implicit
   Crank-Nicolson scheme, a $0.5$ degree latitudinal grid ($56$~km)
   and a two~year timestep.

\begin{table}[h!]
  \begin{center}
    \begin{tabular}{lrll}
      \hline    
      Variable		&	Value	&	&	Description\\
      \hline
 $\;$$\rho_b$		&	$3370$		&kg/m$^3$ &bedrock density \\
 $\;$$\rho_i$		&	$910$		&kg/m$^3$ &ice density \\
 $\;$$\rho_s$		&	$2390$		&kg/m$^3$ &saturated sediment bulk density \\
 $\;$$\phi_s$		&	$22^{\textrm{o}}$&degrees		&angle of internal friction \\
 $\;$$A$			&	$7.7 \times 10^{-29}$ &1/(Pa$^3$~s )	&deformability of ice \\ 
$\;$$D_{o}$		&	$2.5 \times 10^{-14}$&s$^{-1}$ & reference sediment deformation rate \\
 $\;$$m$			&	$1.25$		&	& exponent in sediment stress-strain relationship\\
 $\;$$n$			&	$3$		&	&exponent in Glen's Law \\
 $\;$$h_{\textrm{sed}}$&	$10$		&m	& thickness of sediment layer\\
 $\;$$H_{eq}$		&	$0$			&m & equilibrium height above sea level \\
 $\;$$T_b$		&	$5000$		&years	& bed relaxation time constant \\
 $\;$$u_{o}$		&	$3 \times 10^9$&Pa/s &sediment reference viscosity \\
      \hline
    \end{tabular}
  \end{center}
  \caption{Parameters used for the ice sheet model.}
  \label{Tab:Ice_Parameters}
\end{table}

\subsection{Convolution} \label{sec:convolution} 
In section~\ref{sec:mor-co_2-response}, we investigate the effect of
sinusoidal sea-level on MOR emissions and explain the results in terms
of physical processes. However, results in
figure~\ref{fig:MORC_periodicSL} can be understood by the mathematical
properties of a simplified system: convolving triangular functions,
(approximately our MOR Green's functions) with sine functions (rate of
change of sea level).  This simplified system allows us to define
short period and long period regimes precisely, unlike the physical
argument.  This section details an exact analytical solution for this
simplified system, and shows that the results align with our physical
explanation

We define a triangle function $T(x-d)$ with width $2d$
beginning at $x=0$,where $d$ is the mid-point of the triangle.
Similarly, boxcar functions of width $d$ are denoted by $B(x-d)$,
where $d$ is the mid-point of the boxcar.  Sea level is a sine
function with amplitude $A$ and period $P$. Thus the convolution of
rate-of-change-of-sea-level with the triangle function is:
\begin{align}
T(t)*Ak\cos(kt) = T(t)*\pdiff{A\sin(kt)}{t} \;\;,
\end{align}
where $k = 2\pi/P$. Applying the general convolution property $f*g' = f'*g$,
\begin{align}
T(t)*Ak\cos(kt) = (B(t-d/2)-B(t-3d/2))*A\sin(kt)\;\;.
\end{align}
Repeating this step, and taking the derivative of a Heaviside function (the
edges of the boxcar functions) as a Dirac delta function gives:
\begin{align}
T(t)*Ak\cos(kt) = \big[ \delta(t) - 2\delta(t-d) + \delta(t-2d) \big]*-\frac{A}{k}\cos(kt)\;\;,
\end{align}
which is evaluated to give the percentage change in MOR \cotwo{}
emissions rate
\begin{align}
T(t-d)*Ak\cos(kt) = -\frac{A}{k} \big[\cos(kt) - 2 \cos(k(t - d)) + \cos(k(t - 2d))\big]\;\;.
\end{align}
Integrating this and simplifying gives the cumulative
change in atmospheric \cotwo{}
\begin{align}
C = \frac{AP^2}{\pi^2} \sin^2\left(\frac{d\pi}{P} \right) \sin\left( \frac{2\pi}{P} (d-t) \right)\;\;.\label{eq:triconv}
\end{align}
Equation~\eqref{eq:triconv} is zero for $P=d/n$ where $n \in
\mathbb{Z}_{>0}$.  Therefore the largest sea-level period for which $C=0$
is half the width of the triangle function (\ie{} $P=d$), matching the end of the
small period region in figure~\ref{fig:MORC_periodicSL}.

\begin{figure}[ht]
  \centering
\includegraphics[width=10cm]{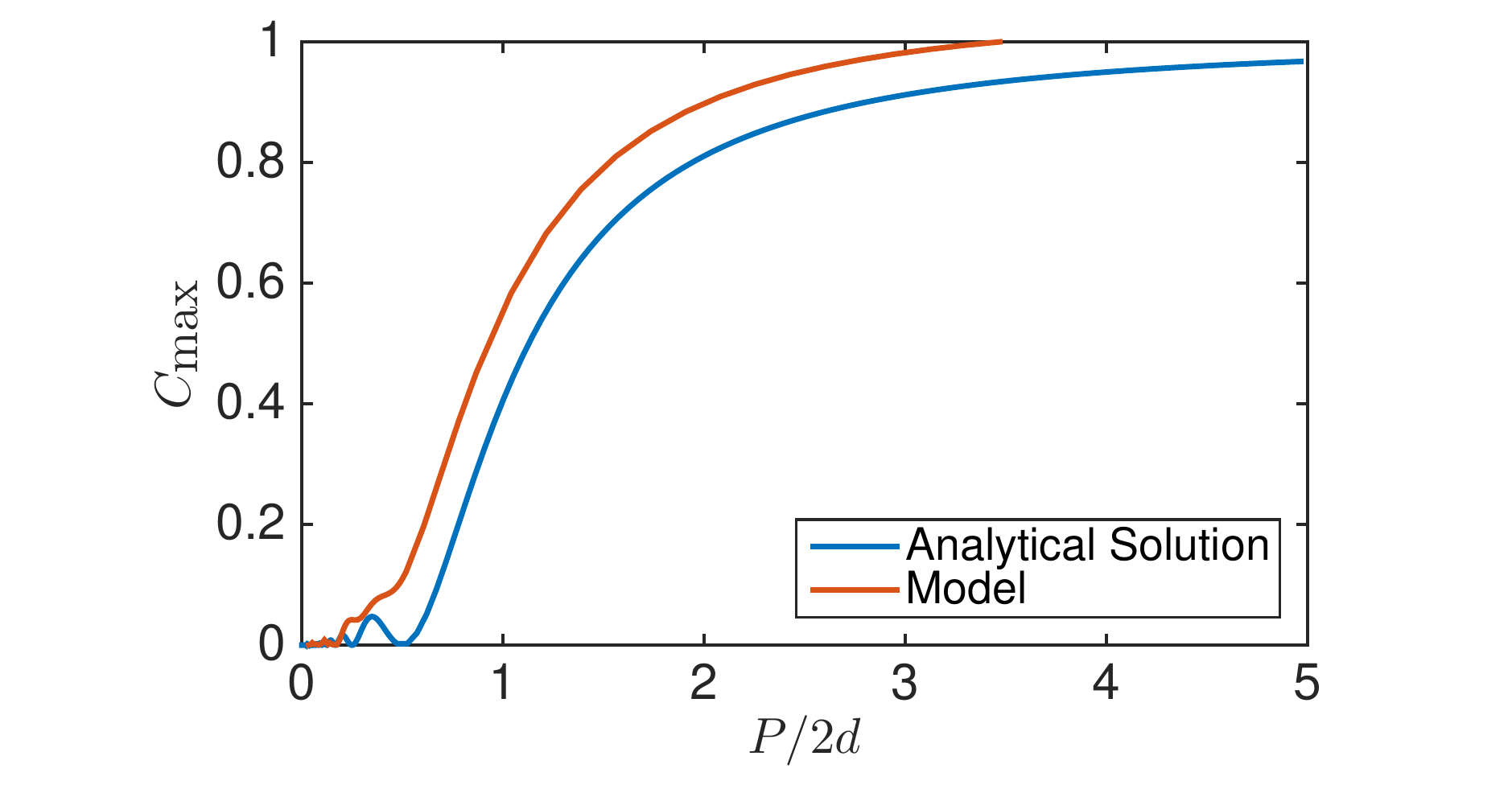}
\caption{Normalised maximum cumulative \cotwo{} emissions for the
  analytical solution of equation~\eqref{eq:C_max_toy}, and the actual
  volcanic Green's function from \ref{fig:MORC_periodicSL}.  The
  x-axis is sinusoidal sea level period divided by the width of the
  Green's function.}
  \label{fig:Conv_Toy}
\end{figure}

The maxima of equation~\eqref{eq:triconv} occur for
$\partial{C}/\partial{t} =0$ and are all equal, thus we can consider
the first maxima as representative, given by:
\begin{align}
 t = \frac{4 d \pi + P\pi}{4 \pi} \label{eq:t4maxima}
\end{align}
Substituting eqn.\eqref{eq:t4maxima} into eqn.~\eqref{eq:triconv} gives an expression
for the maxima of $C$
\begin{align}
 C_{\textrm{max}} = \frac{AP^2}{\pi^2} \sin^2 \left(  \frac{d \pi}{P}  \right)\;\;,\label{eq:C_max_toy}
\end{align}
which has $\lim_{P\to\infty} = d^2$ via the squeeze theorem.
Therefore the maximum amplitude of the change in atmospheric \cotwo{}
concentration is proportional to the width of the triangle function.
Figure~\ref{fig:Conv_Toy} shows that the normalised behaviour of
equation~\eqref{eq:C_max_toy} is similar to our volcanic
Green's functions.

However, MOR volcanism actually converges to a single value,
independent of the Green's function width, therefore this proposed $d^2$ scaling
represents a divergence between our Green's functions of MOR \cotwo{}
emissions and the triangle function approximation.  The difference is
readily explained. As seen in figure~\ref{fig:MORC_periodicSL}a, for
longer MOR lag times, the right-hand-side of the Green's function
triangle becomes increasingly concave compared to a hypothetical,
symmetrical triangle function, causing less emissions than would be
expected from the triangle function approximation.  This difference
is such that the Green's functions all have the same total emissions in
the long sea-level period limit.

Alternatively, explaining in physical terms, recall that
sea-level-driven variable MOR \cotwo{} emissions are caused by the
changing depth of first mantle melting.  A low sea-level means a
deeper depth of first melting, effectively flushing out \cotwo{} that
would occupy mantle at that depth and inserting it into the
atmosphere.  There is a fixed density of \cotwo{} in that mantle, and
thus a finite maximum \cotwo{} mass that a given amplitude of
sea-level change can insert into the atmosphere, regardless of its
period.  Long-period sinusoidal sea level changes $P \gg d$ remove all
the interference effects in upwelling melts arriving at the MOR and
therefore all long-period sea level changes converge towards this same
maximum \cotwo{} mass, regardless of MOR lag.

\subsection{Tuning $C$ Feedback  Strength} \label{sec:tuning-c-feedback}
Equation~\eqref{eq:Carbon_full} for carbon concentration in the
atmosphere $C$ has three terms with corresponding sensitivity factors
$\gT$, $\gMOR$, $\gSAV$ controlling how much $C$ will change in
response to changes in planetary temperature and sea level.  In this
section we run the model with only one \cotwo{} feedback term active,
varying its corresponding sensitivity parameter, for each of the three
feedback terms. The insolation forcing includes obliquity, precession
and eccentricity.  This demonstrates how the model behaves (with
reconstructed insolation) when a particular \cotwo{} feedback is
dominant .

\begin{figure}[ht]
  \centering
\includegraphics[width=12cm]{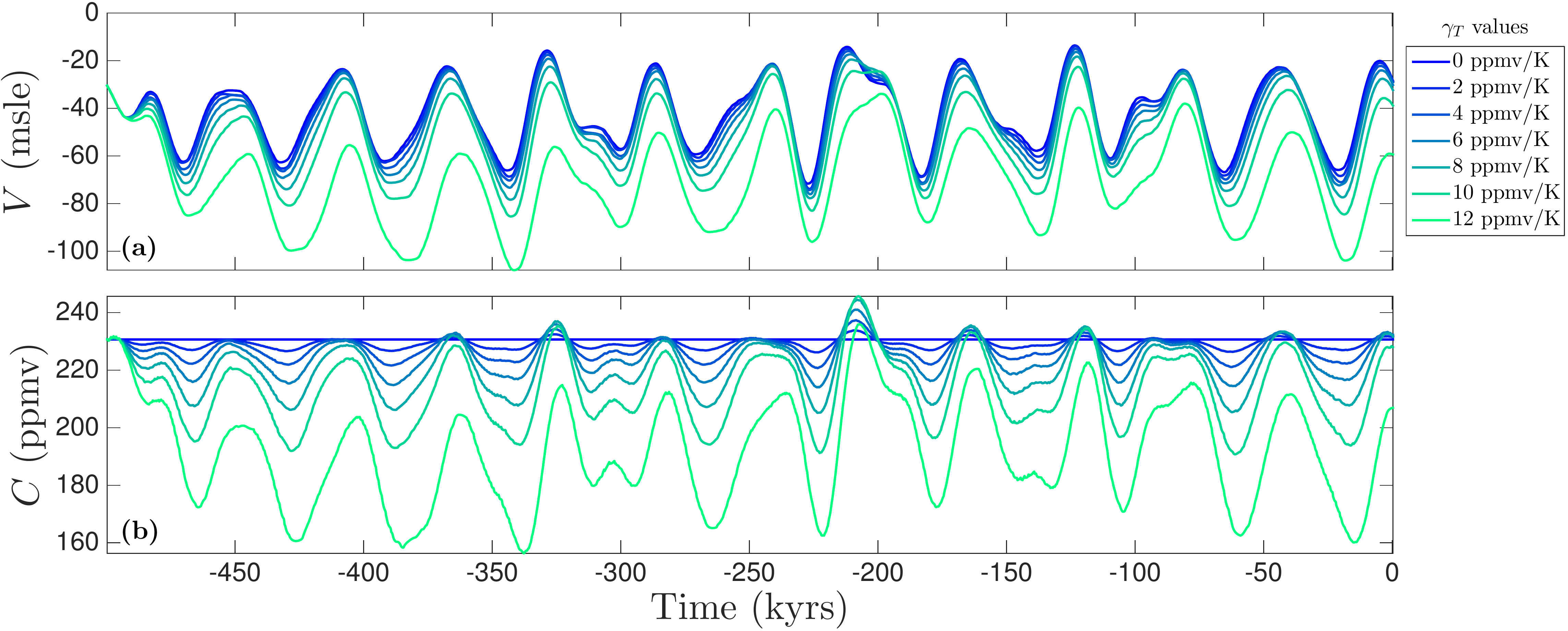}
\caption{ Increasing \cotwo{} temperature sensitivity parameter  $\gT$ with
  volcanic \cotwo{} effects turned off: $\gMOR , \gSAV =0$. Panel
  \textbf{(a)} is ice volume in metres sea level equivalent. Panel \textbf{(b)} is
  \cotwo{} concentration in the atmosphere.}
  \label{fig:g1_ONLY}
\end{figure}
Figure~\ref{fig:g1_ONLY} shows the model with $C$ responding only to
changes in global average temperature: $\gT \neq 0$ and $\gMOR,\gSAV =
0$.  The amplitude of the glacial cycles increases with increasing
positive feedback between $C$ and temperature, until the model enters
a runaway glacial for $\gT > 12$~ppmv/K. 

\begin{figure}[h!]
  \centering
\includegraphics[width=12cm]{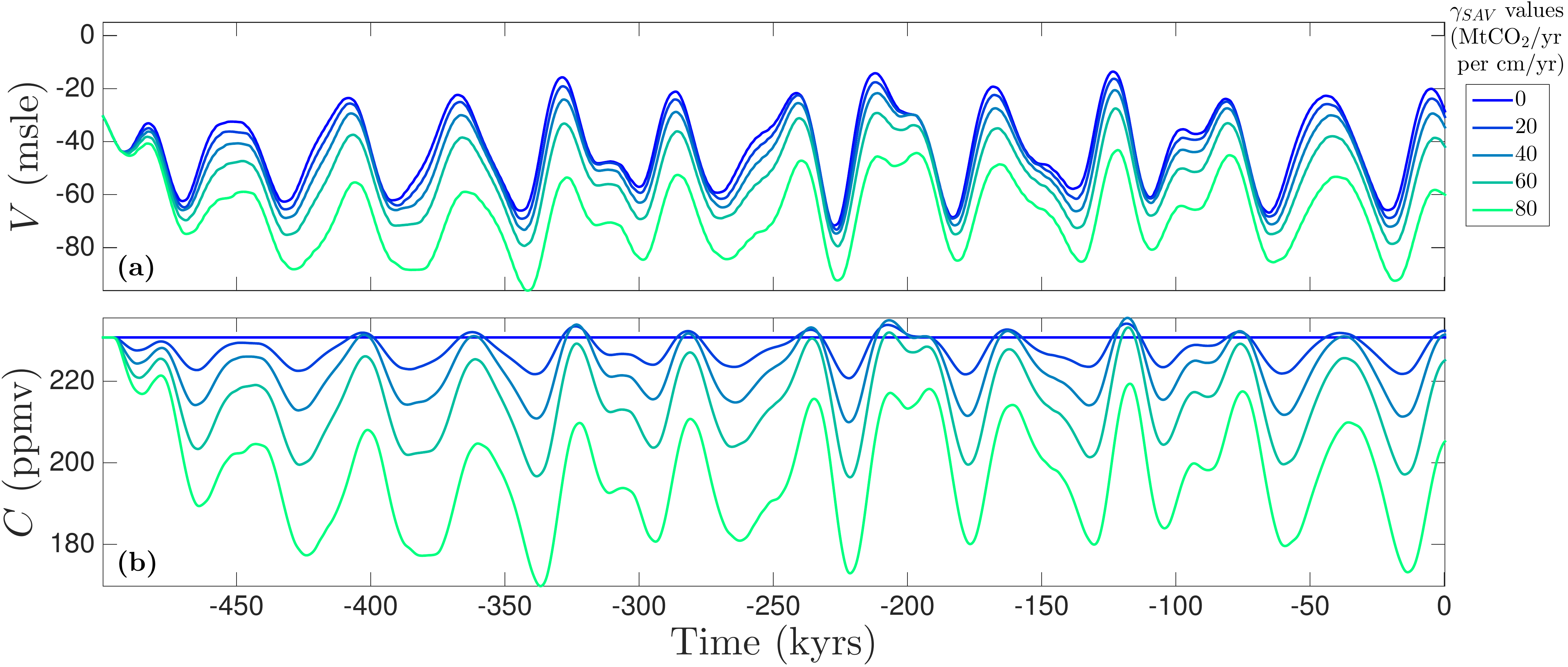}
\caption{ Increasing SAV \cotwo{} sensitivity to sea-level $\gSAV$
  with other \cotwo{} effects turned off: $\gT , \gMOR
  =0$~Mt\cotwo{}/yr per cm/yr. Panel \textbf{(a)} is ice volume in metres sea
  level equivalent. Panel \textbf{(b)} is \cotwo{} concentration in the
  atmosphere. The values in the legend are roughly equivalent to the
  peak increase in SAV \cotwo{} emissions flux caused by a
  deglaciation where $\dot{V}=1$~cm/yr. }
\label{fig:g3_ONLY}
\end{figure}
Figure~\ref{fig:g3_ONLY} has $C$ changing with variable SAV only:
$\gSAV \neq 0$ and $\gT , \gMOR =0$.  SAV \cotwo{} emissions lag the
ice cycle by about 4~kyrs, and cause a runaway glacial for $\gSAV \geq
90$~Mt\cotwo{}/yr per cm/yr.  As expected from the discussion of
short-timescale feedbacks in section~\ref{sec:varying-mor-lag}, both
$V$ and $C$ timeseries in figure~\ref{fig:g3_ONLY} are similar to
figure~\ref{fig:g1_ONLY}, and neither changes the dominant period of
glacial cycles.

\begin{figure}[ht]
  \centering
\includegraphics[width=12cm]{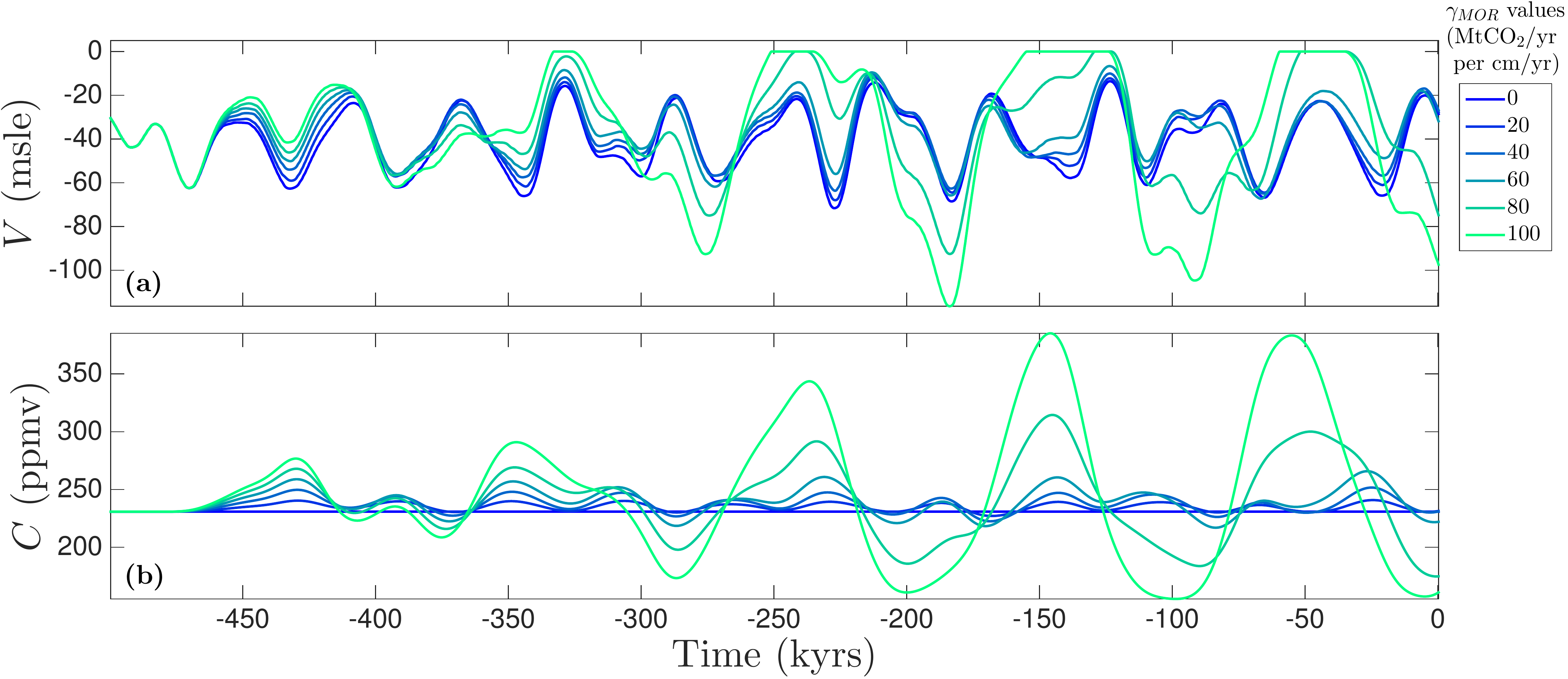}
\caption {Increasing MOR \cotwo{} sensitivity to sea-level $\gMOR$ with other
  \cotwo{} effects turned off: $\gT , \gSAV =0$. Panel \textbf{(a)} is
  ice volume in metres sea level equivalent. Panel \textbf{(b)} is \cotwo{}
  concentration in the atmosphere. The values in the legend
  are roughly equivalent to the peak decrease in MOR \cotwo{}
  emissions flux caused by a deglaciation where $\dot{V}=1$~cm/yr. 
}
  \label{fig:g2_ONLY}
\end{figure}
Figure~\ref{fig:g2_ONLY} has $C$ changing with variable MOR \cotwo{}
emissions only ($40$~kyr MOR lag time): $\gMOR \neq 0$ and
$\gT,\gSAV = 0$.  
As with \gT{} and \gSAV{}, increasing the $C$ sensitivity parameter
\gMOR{} increases the amplitude of $C$ cycles. However, the response
in ice volume is more complex; as $\gMOR$ is increased, different
maxima and minima in $V$ become more extreme or suppressed.  This
occurs because, as in section~\ref{sec:varying-mor-lag}, the MOR-based
variability in $C$ opposes some insolation-driven changes in $V$ (and
reinforces others).  As $\gMOR$ increases further, the model moves
towards glacial cycles at a multiple of the obliquity cycle: for
$\gMOR=80$ Mt\cotwo{}/yr per cm/yr a $\sim$100~kyr oscillation
dominates in the final 300~kyrs of $C$ (fig~\ref{fig:g2_ONLY}b).

These results for varying $C$ terms in eqn~\eqref{eq:Carbon_full} are
consistent with our suggestion that feedbacks need at least a $30$~kyr
lag time to disrupt 40~kyr glacial cycles.  This reinforces the
conclusions from exploring varying MOR lag times in
section~\ref{sec:varying-mor-lag}.
Collectively, the results presented so far imply that variable MOR
\cotwo{} emissions can change both the amplitude and periodicity of
glacial cycles in $C$ and $V$, and the short-timescale carbon feedbacks from
subaerial volcanism and the surface system only change the amplitude
of glacial cycles. However, the increased amplitude of sea level
change causes increased amplitude of cumulative MOR CO$_2$ emissions,
therefore when all \cotwo{} feedbacks are active in the model the short-timescale feedbacks
will affect when the model generates $\sim$100~kyr cycles.

 \clearpage{}

\subsection{Model Response to Noise} \label{sec:model-response-noise}
The main text establishes the behaviour of the model as a 
deterministic system. We showed the model tends toward $40$~kyr
glacials driven by insolation cycles, but transitions to glacials at a
multiple of the $40$~kyr cycle if the MOR \cotwo{} emissions response to
sea-level change is increased.
This section explores the model response to noise, evaluating if the
model's glacial cycles are stable to stochastic forcing of key variables.
Geological records show considerable noise at $\lesssim 500$~year
periods. Would such noise affect model behaviour?

To test this, we add Gaussian white noise to a parameter, randomly
changing its value at each timestep in the model.  The noise
timeseries was created using MATLAB's wgn function. We applied this
model perturbation independently to insolation (change solar constant)
and precipitation rate (change $P_c$).

\begin{figure}[ht]
  \centering
\includegraphics[width=12cm]{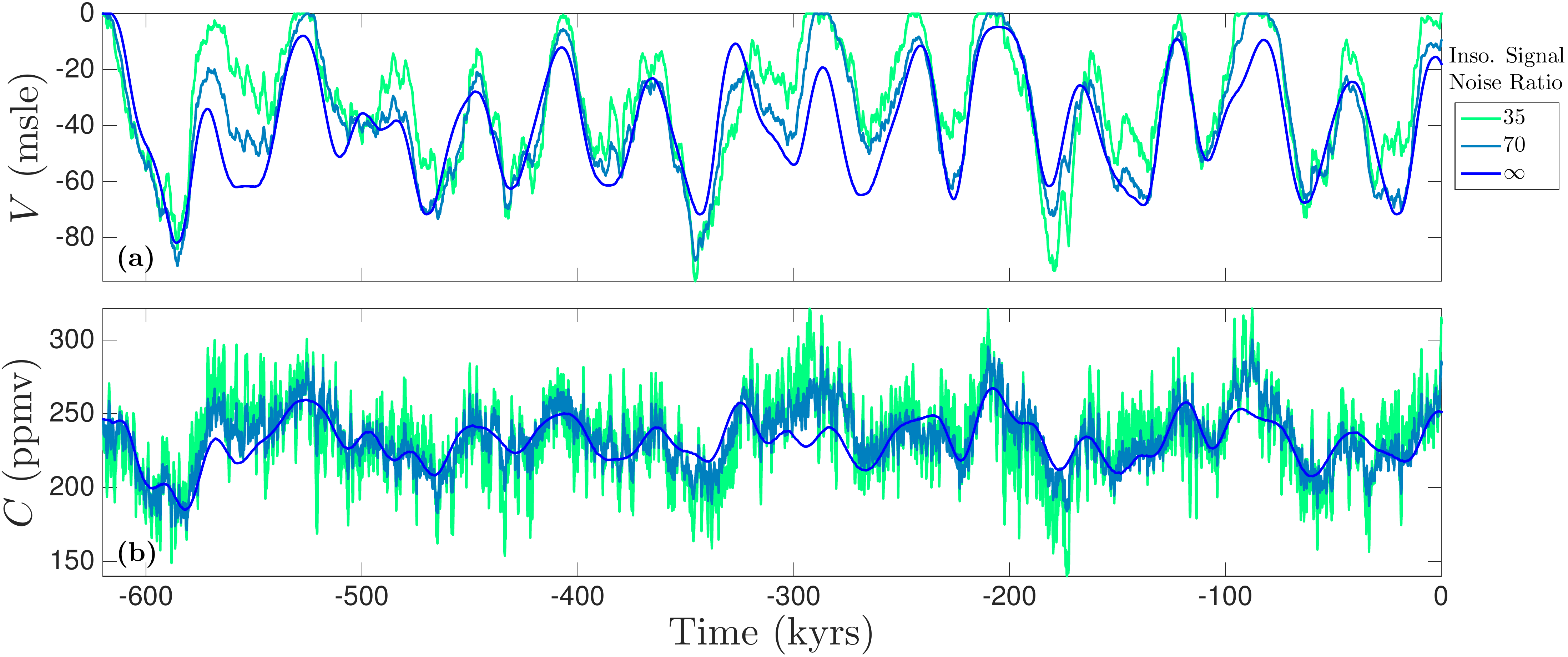}
\includegraphics[width=12cm]{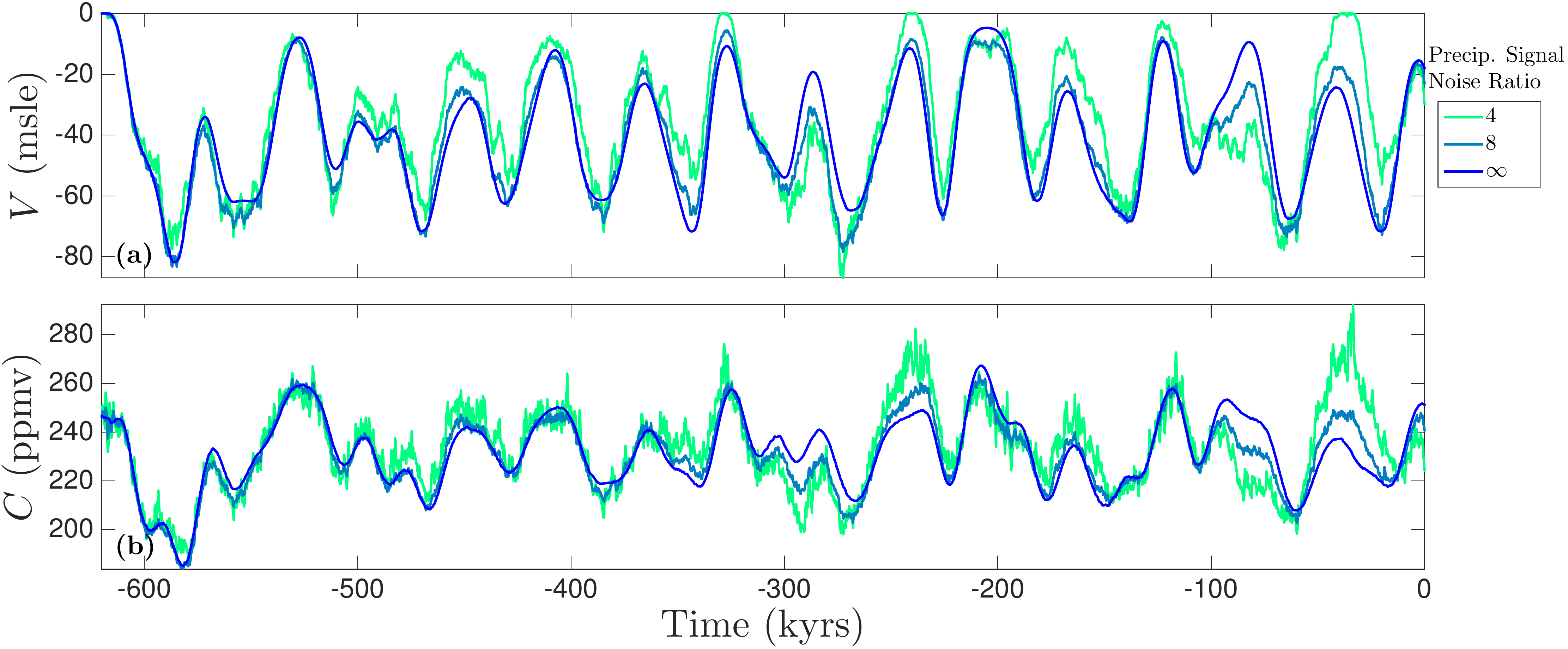}
\caption{Model runs with increasing white noise added to insolation
  (upper pair) and precipitation (lower pair). In each pair the upper
  panel is ice volume in metres sea level equivalent and the lower
  panel is \cotwo{} concentration in the atmosphere.  Noise is
  calculated by MATLAB's wgn function with random seed 400. Signal to
  noise ratio is the standard deviation of the noise divided by
  default value of the perturbed parameter.}
  \label{fig:INSO_RandNoise}
\end{figure}

Figure~\ref{fig:INSO_RandNoise} shows the results of stochastic
forcing in both insolation (fig~\ref{fig:INSO_RandNoise}a,b) and
precipitation (fig~\ref{fig:INSO_RandNoise}c,d).  Both forcings
show high-frequency perturbations in $V$ and $C$, despite the
different physical mechanisms behind the perturbations.  Insolation
forcing changes the shortwave energy flux, thus changing Earth's
temperature and consequently causing both ice sheet growth/retreat and
(via the \gT{} carbon feedback and WLC feedback term) $C$
changes. Precipitation forcing
changes annual snowfall and thus directly affects ice sheet
growth/retreat. The change in ice sheet extent changes surface albedo
and consequently Earth's temperature, leading to $C$
changes.

The stochastic forcings have no significant effect on SAV or MOR
\cotwo{} emissions in the model, as the high-frequency variations in
sea level (with a mean of zero) cancel out by the mechanism explained
in section~\ref{sec:mor-co_2-response}.

Neither insolation nor precipitation forcing significantly
alters the $40$~kyr glacial cycle. These results hold for a range of
random seeds and signal-to-noise ratios.  Therefore high frequency noise
similar to or greater than the geological record does not affect the key
conclusions of this study.
\clearpage{}

\bibliographystyle{abbrvnat} 
\bibliography{Biblio_CVICE}
\end{document}